\documentclass[acmsmall]{acmart}

\AtBeginDocument{%
  }

\setcopyright{acmcopyright}
\copyrightyear{2022}
\acmYear{2022}
\acmDOI{XXXXXXX.XXXXXXX}

\acmJournal{TOIS}
\acmVolume{37}
\acmNumber{4}
\acmArticle{111}
\acmMonth{8}

\usepackage{subfigure}
\usepackage{amsmath}
\usepackage{verbatim}
\usepackage{multirow}

\usepackage{algorithm}
\usepackage{algorithmic}
\usepackage{xcolor}

\usepackage{enumitem}

\begin{document}

\title{Alleviating Video-Length Effect for Micro-video Recommendation}

\author{Yuhan Quan}
\email{quanyh15@tsinghua.org.cn}
\affiliation{
  \institution{Tsinghua University}
  \city{Beijing}
  \country{China}
}

\author{Jingtao Ding}
\authornote{corresponding author}
\email{dingjt15@tsinghua.org.cn}
\affiliation{
  \institution{Tsinghua University}
  \city{Beijing}
  \country{China}
}

\author{Chen Gao}
\email{chgao96@gmail.com}
\affiliation{
  \institution{Tsinghua University}
  \city{Beijing}
  \country{China}
}

\author{Nian Li}
\email{linian21@mails.tsinghua.edu.cn}
\affiliation{
  \institution{Tsinghua University}
  \city{Beijing}
  \country{China}
}

\author{Lingling Yi}
\email{chrisyi@tencent.com}
\affiliation{
  \institution{Tencent}
  \city{Shenzhen}
  \country{China}
}

\author{Depeng Jin}
\email{jindp@tsinghua.edu.cn}
\affiliation{
  \institution{Tsinghua University}
  \city{Beijing}
  \country{China}
}

\author{Yong Li}
\email{liyong07@tsinghua.edu.cn}
\affiliation{
  \institution{Tsinghua University}
  \city{Beijing}
  \country{China}
}

\renewcommand{\shortauthors}{Quan et al.}

\begin{abstract}
  Micro-videos platforms such as TikTok are extremely popular nowadays. One important feature is that users no longer select interested videos from a set, instead they either watch the recommended video or skip to the next one. 
  As a result, the time length of users' watching behavior becomes the most important signal for identifying preferences.
  However, our empirical data analysis has shown a video-length effect that long videos are easier to receive a higher value of average view time, thus adopting such view-time labels for measuring user preferences can easily induce a biased model that favors the longer videos.
  In this paper, we propose a \textbf{V}ideo \textbf{L}ength \textbf{D}ebiasing \textbf{Rec}ommendation~(VLDRec) method to alleviate such an effect for micro-video recommendation. VLDRec designs the data labeling approach and the sample generation module that better capture user preferences in a view-time oriented manner. It further leverages the multi-task learning technique to jointly optimize the above  samples with original biased ones.
  Extensive experiments show that VLDRec can improve the users' view time by 1.81\% and 11.32\% on two real-world datasets, given a recommendation list of a fixed overall video length, compared with the best baseline method. Moreover, VLDRec is also more effective in matching users' interests in terms of the video content.

\end{abstract}



\maketitle

\section{Introduction}
\label{sec: intro}

\begin{figure}
    \centering
    \subfigure[]{\includegraphics[width=.3\textwidth]{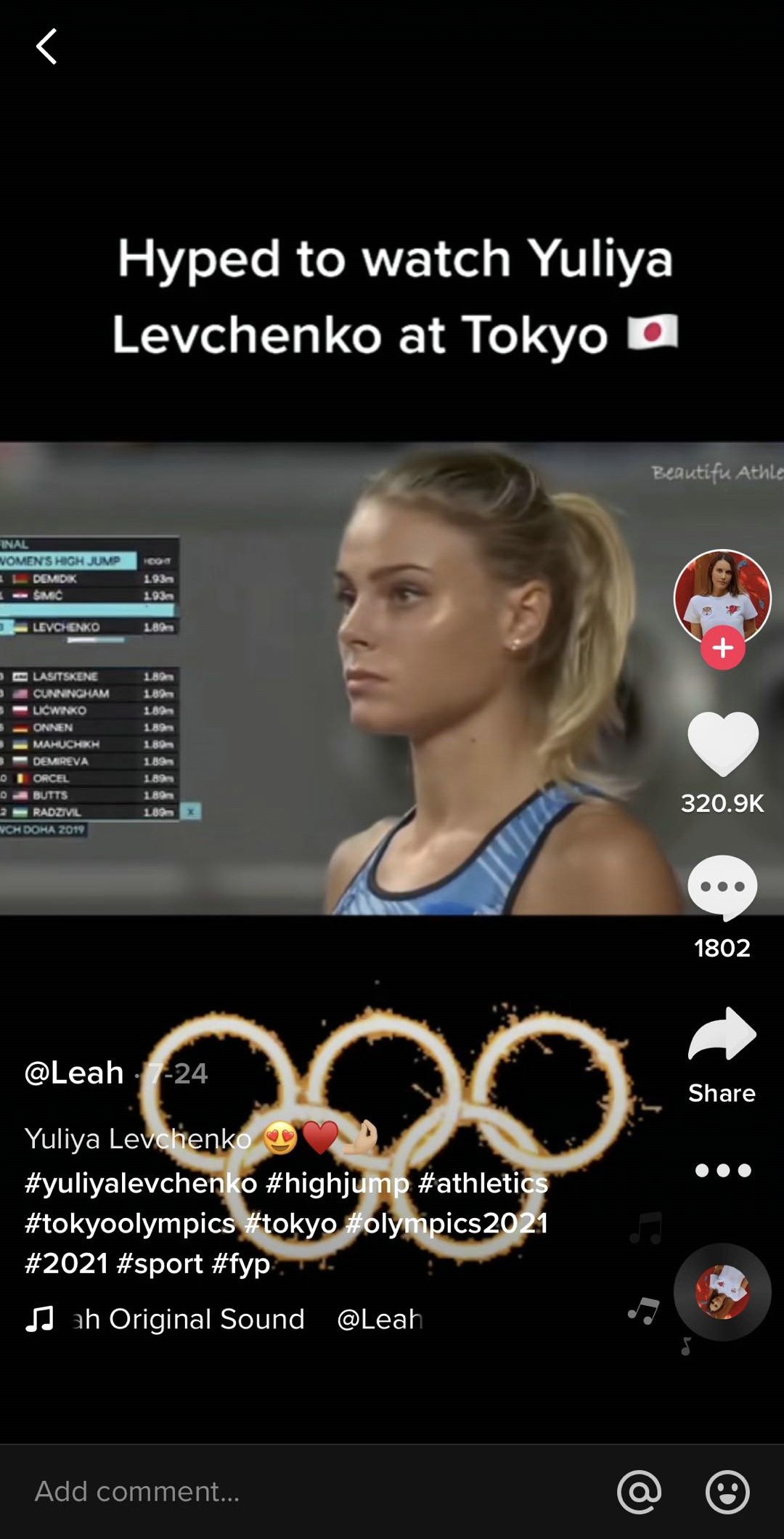}}
    \hspace{.3in}
    \subfigure[]{\includegraphics[width=.3\textwidth]{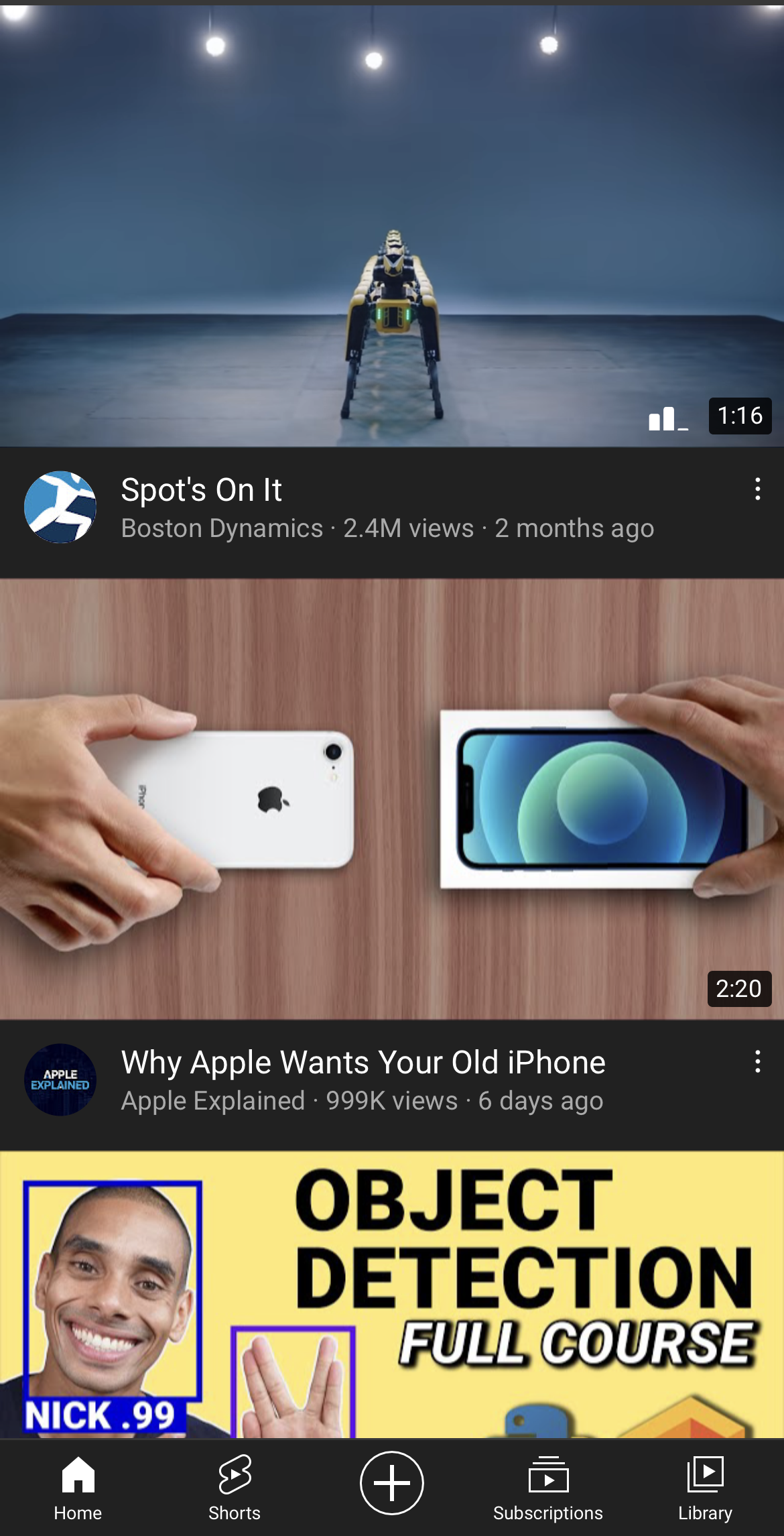}}
    \caption{(a)TikTok~(micro video app), showing only one video at a time. (b)Youtube~(video app), showing multiple videos at a time, \emph{i.e.}, users need to click on the video of interest to watch.}
    \label{fig:comparision}
\end{figure}

Recommender systems, which can provide items that users may be interested in from a large number of item candidates in a personalized way,
are widely deployed nowadays for filtering information or content. 
Recently, with the help of recommender system, micro-video platforms such as TikTok\footnote{https://www.tiktok.com} have swept the world.
In fact, there is a significant gap between the micro-video recommender system with the traditional video platforms, such as YouTube recommendation~\cite{covington2016deep,davidson2010youtube}\footnote{In fact, YouTube's iOS App has also been added a tab for micro-video recommendation recently.}.
In traditional video websites, a user is always shown/recommended a list of videos, and then he/she can select and click one video he/she feels interested in.
As we have mentioned above, the new micro-video platform has completely upgraded the pipeline, of which the most representative one is TikTok.
As shown in Fig.~\ref{fig:comparision}(a)(b), compared with YouTube, the video length is relatively shorter, generally from tens of seconds to several minutes. The brand new paradigm of user-video interaction can be described as: the videos keep continuously playing until the user slides down the screen, after which another video shows in a similar way.

\begin{figure}
    \centering
    \includegraphics[width=.6\textwidth]{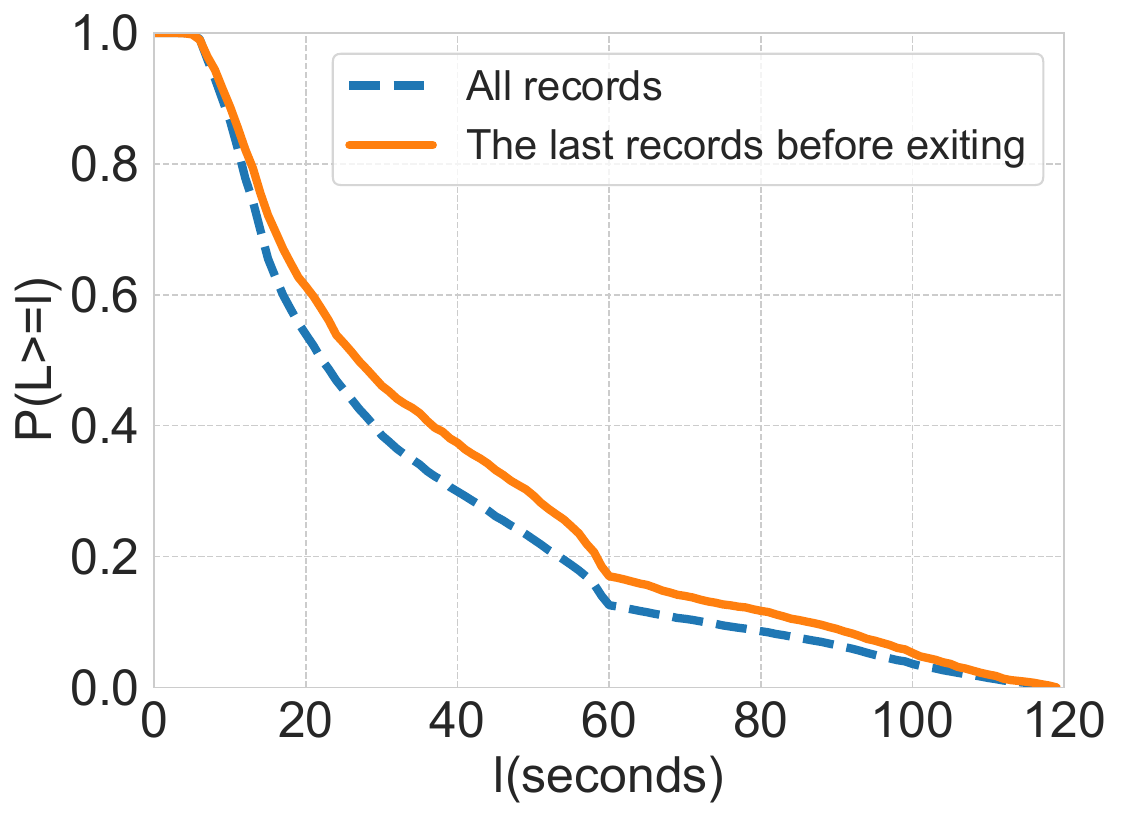}
    \caption{Distribution of video length for all viewing records and the last viewing records before exiting the app~(on wechat dataset).}
    \label{fig:comparision_records}
\end{figure}

This difference in interaction manner has actually caused significant changes in the recommender systems. Specifically, for traditional video recommender systems, the main optimization goal is to increase the user's click-through rate (CTR). If the user clicks on a video, it can be considered that the user is interested in this video and can be seen as a positive signal~\cite{covington2016deep,lu2021multi}.
However, in the brand new micro-video recommender system, users \textit{do not click anymore} since the videos are automatically exposed and played. It means the recommender system can no longer obtain the user's interest through the "click" behavior. Instead, it can only collect a new kind of feedback, the time length that the user watched the video.

An intuitive solution is to first train the model by predicting user-video view time length and then recommending by ranking the predicted results from long ones to short ones.
However, users' interactions with longer-length videos are naturally easier to reach longer view time, which results in a longer video will be more likely to be recommended with such an intuitive solution. 
We call this commonly existing phenomenon in micro-video platforms as \textit{video-length bias}, where the longer-length videos are particularly favored in the above view-time oriented recommendation scenario. 
Although it seems acceptable for the platform to recommend longer videos and receive a generally higher value of total view-time, the user engagement might be harmed due to the following two reasons. 
First, long videos are easier to cause user fatigue, which has been confirmed by our empirical data analysis of users' exit behavior in a micro-video platform, Wechat\footnote{\url{https://www.wechat.com/en}} Channels. As shown in Fig.~\ref{fig:comparision_records}, the last watched video before a user exits the platform tends to be longer than other ones. 
Second, this unfair advantage of long videos during the learning process possibly includes those undesirable or low-quality videos in the major group, which is known as bias amplification and further hurts recommendation accuracy~\cite{wang2021deconfounded}.
Therefore, in this paper, we focus on alleviating the above video-length bias for micro-video recommendation.

Similar to previous studies of popularity bias and position bias in recommender systems~\cite{abdollahpouri2020connection,gao2022causal}, the collected data with video-length bias also exhibits the distorted user preferences, where longer view time might be caused by longer video length instead of user satisfaction.
However, this new research problem faces two unique challenges that rule out the off-the-shelf debias solutions in other problems like popularity bias and position bias. \textbf{First, the continuous characteristics of both video length and view time largely increase the modeling difficulty of unbiased learning\footnote{Video length and view time both have a time unit of millisecond, and they are almost continuous.}.} On the one hand, characterizing the effect of a continuous attribute, \emph{i.e.}, video length in our case, generally requires a suitable quantization to avoid data sparsity. On the other hand, learning with the continuous view-time label may suffer from the extreme values that 
worsen the variance issue of previous unbiased learning methods like inverse propensity scoring~(IPS)~\cite{joachims2017unbiased,ai2018unbiased}.
\textbf{Second, the complex relation between video length and view time makes it non-trivial to identify true user preferences.}
Generally, in previous studies, the bias factor either impacts the observation probability of a specific item, as in cases of popularity bias or position bias, or correlates with users' intrinsic interests among different item groups, as in the phenomenon of bias amplification.
However, in our case of video-length bias, the bias factor, \emph{i.e.}, the video length, directly impacts the measurement of the preference indicator, \emph{i.e.}, the view time, making it challenging to define a proper label that can represent user preferences in view-time oriented recommendation scenarios.

In this paper, we propose a micro-video recommendation approach named \textbf{V}ideo \textbf{L}ength \textbf{D}ebiasing \textbf{Rec}ommendation~(short for VLDRec) to alleviate the video-length effect.
To overcome the first challenge of modeling continuous video-length effect and continuous view-time labels, we first devise a suitable video grouping based on video length, which is motivated by an important data observation that videos with similar time lengths have a similar distribution of completion rate. Then we adopt a simple workaround to continuous label issues by following a learning-to-rank modeling framework.
In terms of the second challenge in handling the complex relation between video length and view time, we leverage two bias-alleviating data labeling approaches that can better capture users' real preferences regardless of video length. Combined with a length-conditioned sample generation module and a multi-task user preference learning strategy, our proposed VLDRec can achieve an undistorted model training process with the collected view data under a severe video-length bias. Moreover, besides the model training, to ensure fair comparisons regardless of the observed biased video length effect, VLDRec further incorporates a simple but undistorted Top-$T$ metric for evaluating model performance in the micro-video recommendation.

To summarize, the major contributions of this work are as follows:
\begin{itemize}[leftmargin=*]
    \item Different from traditional recommender systems, we approach the brand new problem in the micro-video recommendation, of which the biased video length effect widely exists and may worsen the recommendation performance.
    \item  Motivated by empirical observations regarding the relationship between video length and view time, we propose a novel and general micro-video recommendation method including three parts of bias-alleviating data labeling, length-conditioned sample generation and multi-task user preference learning, which further incorporates a length-invariant Top-$T$ evaluation metric to alleviate the video length effect in both modeling training and evaluation.
    \item We conduct extensive experiments on both public and industrial datasets, and the experimental results demonstrate the effectiveness of our proposed VLDRec method in terms of both longer view time and higher user interest fitness.

\end{itemize}

The rest of this paper is as follows. We discuss the related works in Section~\ref{sec:related works}. In Section~\ref{sec: problem}, we present the motivation from real-world data and formulate the research problem. We then introduce our proposed VLDRec method in detail in Section~\ref{sec: framework}. We conduct experiments in Section~\ref{sec: experiment}. Finally, in Section~\ref{sec: conclusion}, we conclude this paper and discuss the future works.

\section{Related works}
\label{sec:related works}

\subsection{Debias in Recommendation}
In the recommender system, bias has a variety of sources~\cite{chen2021bias}. Common bias includes popularity bias~\cite{mena2021popularity} which caused by popular items being more exposed, selection bias~\cite{pal2012exploring} which caused by the fact that users only select items they are interested in, exposure bias~\cite{ding2020improving} which caused by the fact that users can only interact with exposed items, etc. To alleviate these biases, Inverse propensity scoring~(IPS)~\cite{joachims2017unbiased, schnabel2016recommendations, ai2018unbiased, qin2020attribute} and its improved methods~\cite{bottou2013counterfactual, gruson2019offline} are the most commonly used debiasing methods, which main idea is lowering the weight of the items which have advantages in the recommended results. Therefore, the model will be less influenced by biased instances. In addition, for the bias caused by missing or noisy data, it is usually solved by the relabel method, including heuristic~\cite{steck2010training} or model-based methods~\cite{wang2019doubly,quan2023robust}.  Also, IPS and relabel methods can be combined~\cite{vardasbi2020inverse}. Besides, there are some recently proposed methods, such as disentangling method~\cite{ma2019learning, wang2020disentangled, zheng2021disentangling}, counterfactual method~\cite{wei2021model} or causal graph based method~\cite{zhang2021causal, wang2021clicks,xu2022causal}. These methods alleviate the bias by designing unbiased training targets or instances, or adjusting the prediction results of the model based on mathematical principles. These methods all have some effect on the biases they deal with.
Although there has been a lot of related work on debiasing recommender system, most of these bias will only affect the display of recommendation results. However, this paper focuses on a different problem where the video length directly affects the label of the samples. 
Another angle of analyzing possible impact of video-length effect is fairness~\cite{li2022fairness}, as the biased modeling among videos of different video-length can induce unfair exposure favoring long videos~\cite{li2021user,wang2022make}.

Moreover, we notice that there are some concurrent works that proposes a debiasing recommendation method to handle duration bias in micro-video recommender systems~\cite{zhan2022deconfounding, zheng2022dvr}. Although this duration bias problem is similar to the aforementioned video length effect, we propose to alleviate the biased effect by following the idea of regularization~\cite{chen2021bias}, \emph{i.e.}, learning intrinsic user preferences from less biased feedback data, while above concurrent work~\cite{zhan2022deconfounding} follows the idea of causal inference and another work~\cite{zheng2022dvr} propose a new unbiased prediction objective remove bias and optimize this objective by adversarial learning.

In general, compared to the related work on debiasing in recommender systems, our work follows a data-driven paradigm, designing bias alleviating method motivated by empirical observations instead of directly relying on theories like causal inference, and is optimized for user viewing time goals, which is a key performance indicator in micro-video recommendations. Besides learning method, we further incorporate a length-invariant Top-$T$ evaluation metric to alleviate the video length effect in both model training and evaluation.

\subsection{Video Recommendation}
Video recommendation is an important topic in the recommender system. In terms of content and user interface, there are some differences between video recommendation and recommender systems suitable for e-commerce, news or other scenarios. On the one hand, due to the large amounts of multimedia information and features in videos, some works try to effectively utilize the visual or multimedia features in videos~\cite{yi2022multi, chen2017attentive}. These works focus on how to extract effective features or mix multimodal features. On the other hand, in the scenario of video recommendation, some user behaviors and patterns are different from other scenarios, such as accidentally watching behavior~\cite{wei2020graph}, dynamic interest~\cite{lu2021multi, lu2018deep}, purchase intention through disseminating micro-videos~\cite{lei2021semi}, multimodal interests~\cite{wang2021dualgnn,gao2023survey} and other implicit and explicit feedback~\cite{vallet2011effects, ding2020simplify}. 

Although there have been various works related to video recommendation, our work does not focus on how to extract and utilize the multimedia features of videos, or user behaviors. We mainly focus on the biased video length effect in recommendation due to changes in the user interface in micro-video platforms and try to alleviate this bias as much as possible, so the recommendation model can accurately identify user preferences.

\section{Problem Formulation and Data Observation}
\label{sec: problem}
We first give the problem formulation of the micro-video recommendation that mainly aims to maximize users' view time.
Then we conduct a preliminary analysis on collected user-video interaction data, which motivates our design of the proposed VLDRec method.

\subsection{Problem Formulation}\label{sec: problem-def}
Let $U = \{u_1, u_2,..., u_M\}$ and $V = \{v_1, v_2,...v_N\}$ denote a set of $M$ users and a set $N$ videos, respectively. The video length of $v_j$ is represented as $l_j$. 
For each user $u_i\in U$, given the set of his/her historical interactions $S_{u_i}$, each $(v_k, t_{ik})\in S_{u_i}$ representing $u_i$ has watched $v_k$ with a length of $t_{ik}$, the target is to recommend a new video $v_j\in \{v_k|v_k \in V \land v_k \notin S_{u_i}\}$ with the highest view time $\hat{t}_{ij}$.
Specifically, we formulate this task as a learn-to-rank problem by learning a scoring function $f(u_i, v_j)$ to indicate $u_i$'s preference on $v_j$, which measures $u_i$'s willingness to watch $v_j$ with a long time $t_{ij}$ instead of skipping to a next recommended video. When user $u_i$ watches a video $v_k$ with time  $t_{ik}$, the triplet $(u_i, v_k, t_{ik})$ is defined as a sample. If $p_{ik}\ge1$, then the sample is a completed sample.

The notations we use and their descriptions are shown in Table~\ref{tab:notation}.

\begin{table}[]
\centering
    \caption{The description of notations.}
    \begin{tabular}{c|l}
    \toprule
        Notations & Description \\
    \midrule
        $U$, $V$ & Set of users and videos. \\
        $M$, $N$ & The size of user sets and video sets. \\
        $u_i$, $v_j$ & The specific user and video. \\
        $\textbf{u}_i$, $\textbf{v}_j$ & The embedding of specific user and video. \\
        $l_j$ & The length of video $j$\\
        $t_{ij}$ & The view time of video $j$ watched by user $i$ \\
        $p_{ij}$ & The play progress of video $j$ watched by user $i$ \\
        $S_{u_i}$ & The set of videos watched by user $i$ \\
        $g_v$ & The v-th group of video sets \\
        $v^{un}_k$ & Video $k$ sampled from the group of positive instance \\
        $\tau$ & Threshold for pointwise hard labeling \\
        $\alpha$, $\beta$ & Hyper-parameter of multi-task learning and sampling \\
        $\theta$ & Model parameter \\ 
        $ L_{BPR}$ & The BPR loss function \\ 
        $ f $ & The preference score prediction function for users and videos \\
        $ \hat{\Psi}_{uni} $ & The uniform sampler \\
        $ S^{-}_{u_i,v_j} $ & The set of videos watched by user $i$ with shorter view time compared with $(u_i, v_j)$  \\
        
    \bottomrule
    \end{tabular}
 \label{tab:notation}
\end{table}

\subsection{Data Observation}
\label{sec:data_observation}
\begin{figure}
    \centering
    \subfigure[Kuaishou]{\includegraphics[width=.46\textwidth]{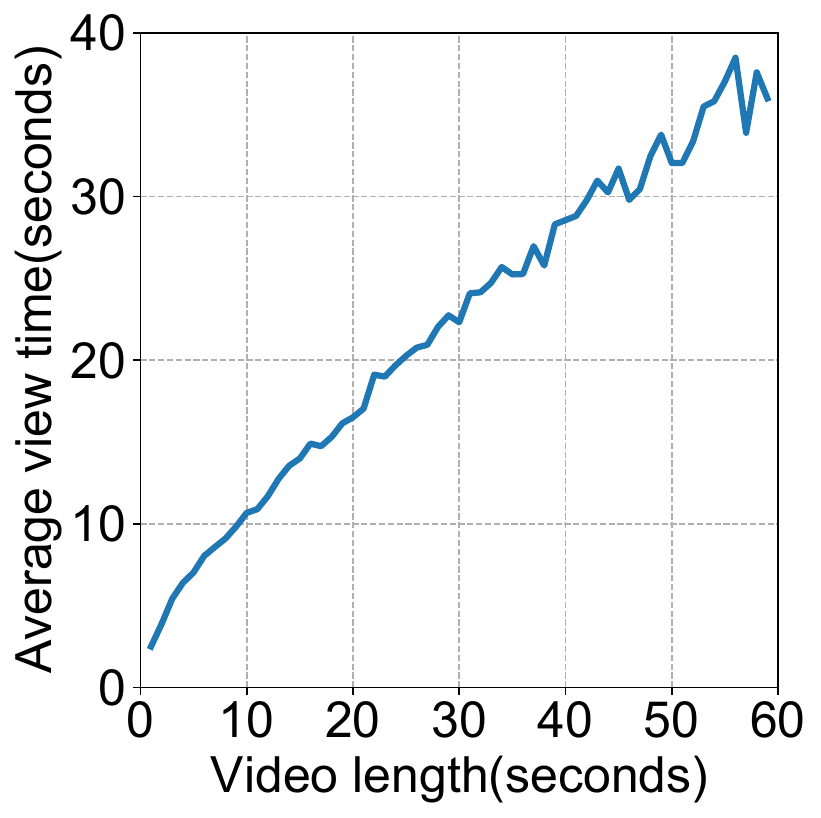}}
    \subfigure[Wechat]{\includegraphics[width=.46\textwidth]{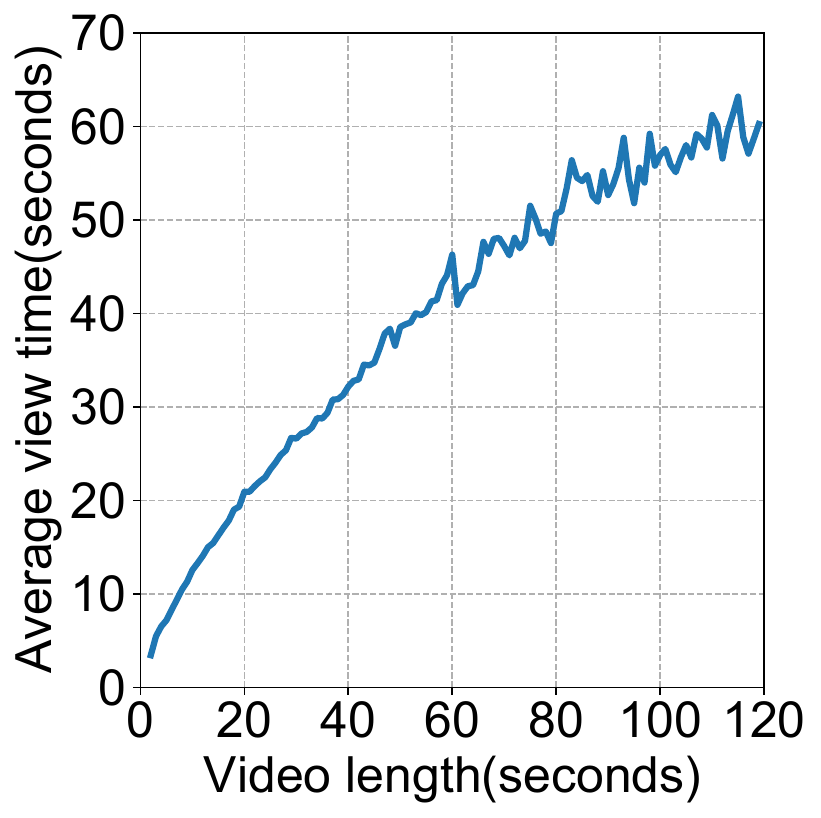}}
    \caption{Average view time of videos of different length.}
    \label{fig:average_view_time}
\end{figure}

\begin{figure}
    \centering
    \subfigure[Kuaishou]{\includegraphics[width=.46\textwidth]{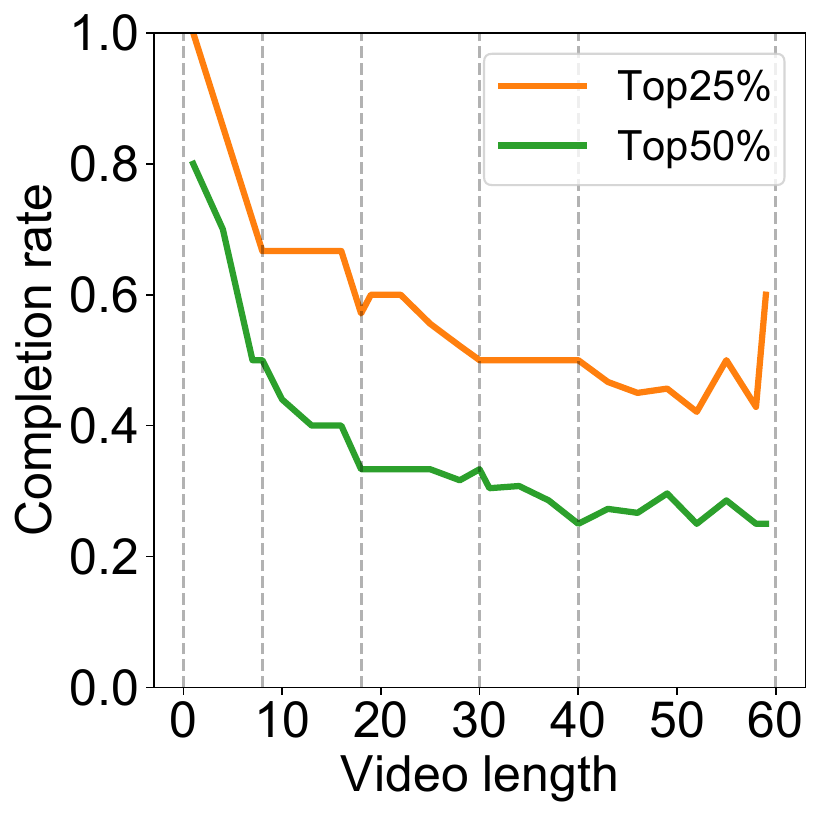}}
    \subfigure[Wechat]{\includegraphics[width=.46\textwidth]{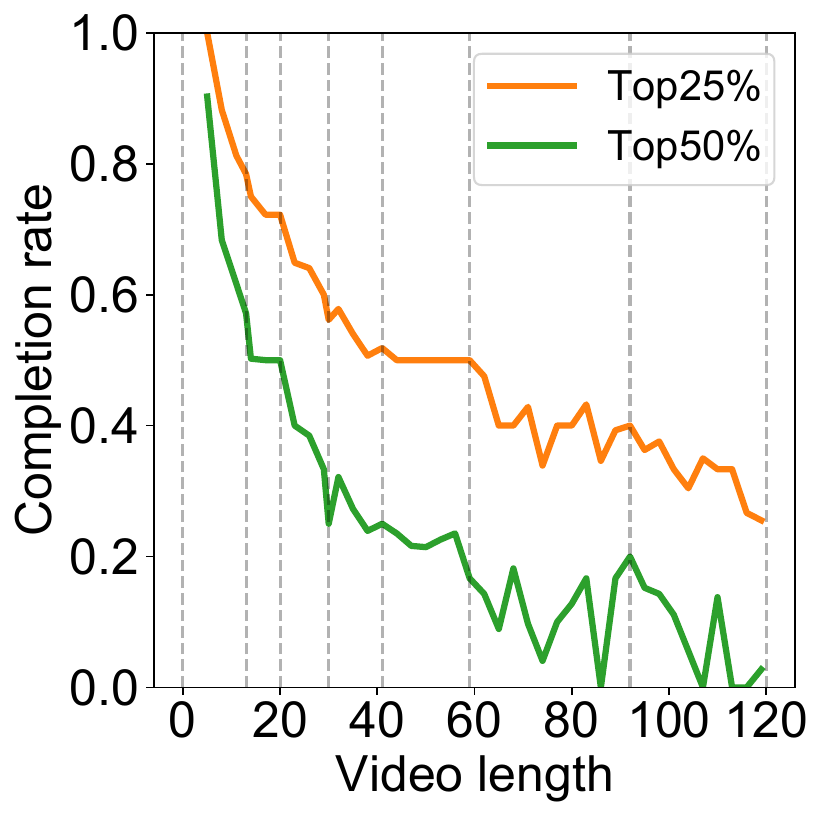}}
    \caption{The relationship between completion rate and video length~(aggregated statistics every one second).}
    \label{fig:dataset_group}
\end{figure}

In order to investigate and understand user behaviors in micro-video recommendation scenarios, we choose two large-scale datasets collected from two leading micro-video platforms, \emph{i.e.}, Kuaishou\footnote{\url{https://www.kuaishou.com/en}} and Wechat Channels\footnote{\url{https://channels.weixin.qq.com}.} 
We leave the detailed descriptions of the above two datasets in Section~\ref{sec: experiment}.  
Here we focus on two indicators of user engagement on recommended micro-videos, \emph{i.e.}, view time and completion rate~\cite{mmoe}.
The former is intuitive, but we will highlight its potential bias in favoring longer videos by illustrating its relationship with video length in Fig.~\ref{fig:average_view_time}(a) and (b).
Comparatively, the latter counts the proportion of completed plays for a specific video, which is normalized into $[0,1]$ by considering both view time and video length.
Mathematically, the completion rate of $v_j$ is equal to the number of completed samples~(\emph{i.e.}, $t_{\bullet j} \ge l_j$) divided by the total number of samples corresponding to the video, which is as follows,
\begin{equation}
    Completion\_rate_v = \frac{\#completed\_samples_v}{\#All\_samples_v}.
\end{equation}
In Fig.~\ref{fig:dataset_group}, by analyzing the distribution of completion rate~($p25$, $p50$ values) for micro-videos with the same video length, we demonstrate that this metric is much fair when comparing users' preference among micro-videos with similar video length.

Specifically, our preliminary analysis has the following two key observations.
\begin{itemize}[leftmargin=*]
    \item \textbf{Long videos are much easier to receive a higher value of average view time.} As shown in Fig.~\ref{fig:average_view_time}(a) and (b), real-world data in both KuaiShou and Wechat follow a similar pattern that the average view time is nearly linear to the increasing video length. Consequently, view time is a biased metric when measuring a user's preference on a micro-video without considering the specific video length. This can lead to severe performance degradation when we attempt to capture user preference with traditional approaches like regression.
    \item \textbf{Videos with similar time lengths have a similar distribution of completion rate.} As shown in Fig.~\ref{fig:dataset_group}(a)~(KuaiShou), by illustrating the $p75$~(the third quartile, \emph{i.e.}, ranked at the top 25\%) and $p50$~(the median) distribution values of micro-videos ranging from 1s to 59s, we can roughly divide them into five groups according to video length, with the flat curve within each group representing a similar distribution pattern. Specifically, these five groups are $[1-8s, 8-18s, 19s-30s, 31s-40s, 41-59s]$. For example, we can observe that micro-videos ranging from 30s to 40s tend to have a completion rate of about $0.5$~($p75$). Similarly for Wechat Channel dataset~(Fig.~\ref{fig:dataset_group}(b)), the suitable grouping is $[0-13s, 14-20s, 21-30s, 31-41s, 42-59s, 60-92s, 93-120s]$. In a word, to capture unbiased user preference from their viewed micro-videos, one needs to extract pairwise ranking relations conditioned on the video length.
   
\end{itemize}

In short, there exists a bias of video-length in video-watching behaviors.

\section{Method}
\label{sec: framework}
\begin{figure*}
    \centering
    \includegraphics[width=.95\textwidth]{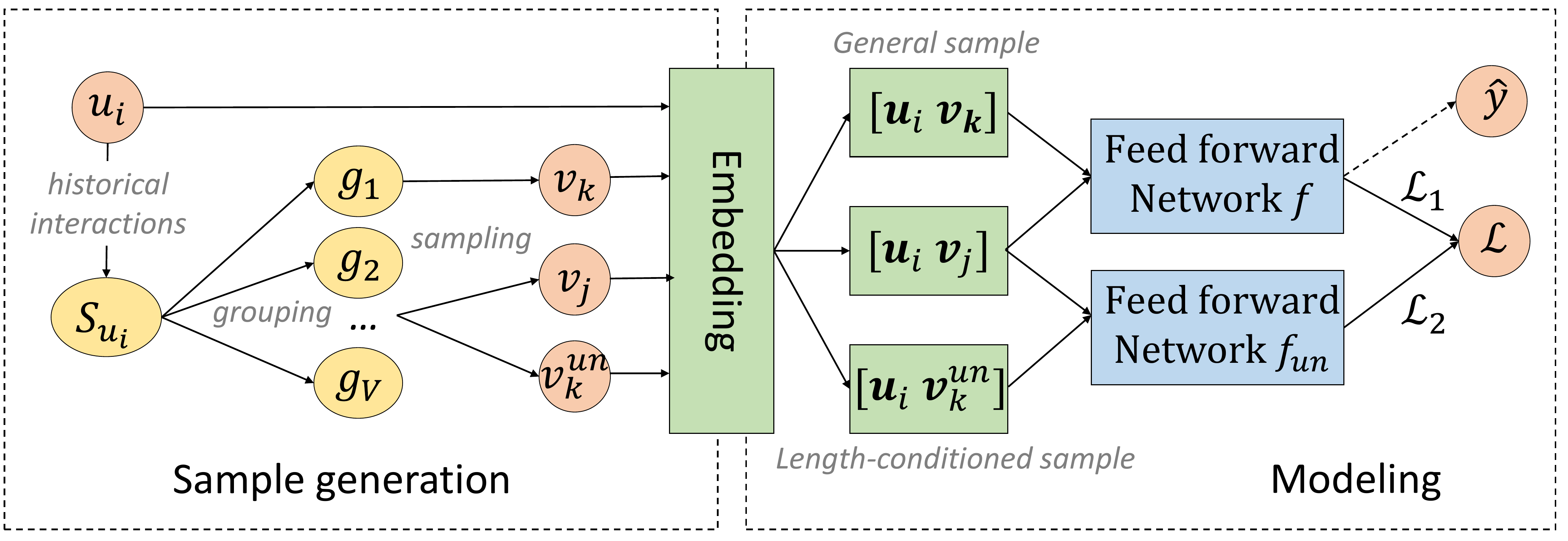}
    \caption{Overall framework of VLDRec}
    \label{fig:overall framework}
\end{figure*}

The proposed VLDRec method models user preferences in a view-time oriented learning-to-rank manner, where two specific labeling approaches are designed to alleviate the video-length effect that can distort the learned user preference. Motivated by previous analysis regarding to the biased preference signal of view time, VLDRec further integrates a length-conditioned sample generation module, which is jointly optimized via multi-task learning. Moreover, to better evaluate micro-video recommendation models, VLDRec adopts a simple but effective Top-$T$ evaluation metric that can make fair comparisons regardless of the observed biased video length effect. The overall framework of VLDRec is shown in fig~\ref{fig:overall framework}.

\subsection{Labeling Approach for View-time Oriented Learning-to-rank}
\label{sec:ltr-approach}

For the micro-video recommendation problem defined in Section~\ref{sec: problem-def}, we adopt a learning-to-rank approach that trains a model to recommend micro-videos with a large probability of generating user engagements, \emph{i.e.}, views.

Generally, to learn recommender models from implicit feedback, Rendle \textit{et al.}~\cite{rendle2012bpr} proposed the Bayesian Personalized Ranking~(BPR) method, which assumes that a positive instance should be predicted with a much higher score over the negative one. Based on BPR, the training objective of the recommender model can be formulated as minimizing the following loss function:

\begin{equation}
\begin{aligned}
    & \! L_{BPR} \!=\! \!\!\sum_{(u_i, v_j)\in S_{u_i}}\!\!\!\! -\ln \sigma(f(u_i, v_j) - f(u_i, v_k)), \\
    & \text{where } v_k \sim \hat{\Psi}_{uni}(S^{-}_{u_i,v_j}).\!\!
\end{aligned}
\label{Lr}
\end{equation}
For each user $u_i$, the predicted preference score on micro-videos is denoted as $f(u_i, \bullet)$. The negative instance $v_k$ is generated by a uniform sampler $\hat{\Psi}_{uni}$ that takes the candidate set $S^{-}_{u_i,v_j}$ as an input, while the positive instance $i$ is randomly chosen from ground truth set $S_{u_i}$. Minimizing $L_{BPR}$ is equivalent to maximizing the margin between $f(u_i, v_j)$ and $f(u_i, v_k)$, which encourages the recommender to learn the pairwise ranking relation of user preference between $v_j$ and $v_k$.

Therefore, an intuitive solution for solving the micro-video recommendation problem is directly putting the instances with longer view time ahead of those with a shorter length. Specifically, when $S^{-}_{u_i,v_j}$ is the set of past instances with shorter view time than current positive instance $(u_i, v_j)$, \emph{i.e.}, $\{(u_i,v_k)|t_{ik}<t_{ij}\}$, the trained model is able to recommend micro-videos that may be watched longer by users.
However, as we discussed in the preliminary analysis, the indicator of view time suffers from the biased video length effect by favoring longer videos, which may cause the distortion of learned user preference.

In our proposed VLDRec model, we instead choose to measure user preference based on another indicator of play progress, \emph{i.e.}, $p_{ij}=t_{ij}/l_j$.
Specifically, we adopt two different definitions of progress-based labels.
The first is pointwise hard labeling depending on whether $p_{ij}$ exceeds a certain threshold $\tau$. Motivated by the previous observation that videos with similar time lengths have a similar distribution of completion rate, we set $\tau$ as $\tau(l_j)$, relevant to the video length $l_j$. According to our analysis in Fig.~\ref{fig:dataset_group}, the micro-videos can be divided into several groups based on their length, \emph{i.e.}, $\{g_1, g_2,...g_V\}$, where $V$ is 5~(KuaiShou) and 7~(WeChat Channel), respectively. 
For these two datasets, we decided the boundary value of video length in each group by checking whether at least one of Top-25\% and Top-50\% curves in Fig.~\ref{fig:dataset_group} would go through a significant change (\emph{i.e.}, from decreasing to non-decreasing or from increasing to non-increasing) at this point compared with neighboring points This operation is performed manually.
Thus we allow videos within the same group sharing the same $\tau=\tau(l_j)=\tau(g)$. Without loss of generality, $\tau(g)$ for each group is set as $p80$ value of the play progress distribution. In other words, the top 20\% samples ranked by their play progress values are considered as positive instances under this pointwise hard labeling.

Besides, we also use another pairwise margin-based labeling. It sets a constraint that there should exist a margin $\epsilon$ between the progress values of a positive instance $(u_i,v_j)$ and a negative instance $(u_i,v_k)$, \emph{i.e.}, satisfying $p_{ij}-p_{ik}>\epsilon$. 
Compared with the above pointwise hard labeling, this pairwise labeling focuses on comparing two samples belonging to the same user, which is more friendly to users with rich interaction history.
Therefore, we leverage the advantages of both two approaches by switching between them, which is controlled by hyper-parameters $\beta$.

\begin{figure}
    \centering
    \subfigure[Only based on view time]{\includegraphics[width=.49\textwidth]{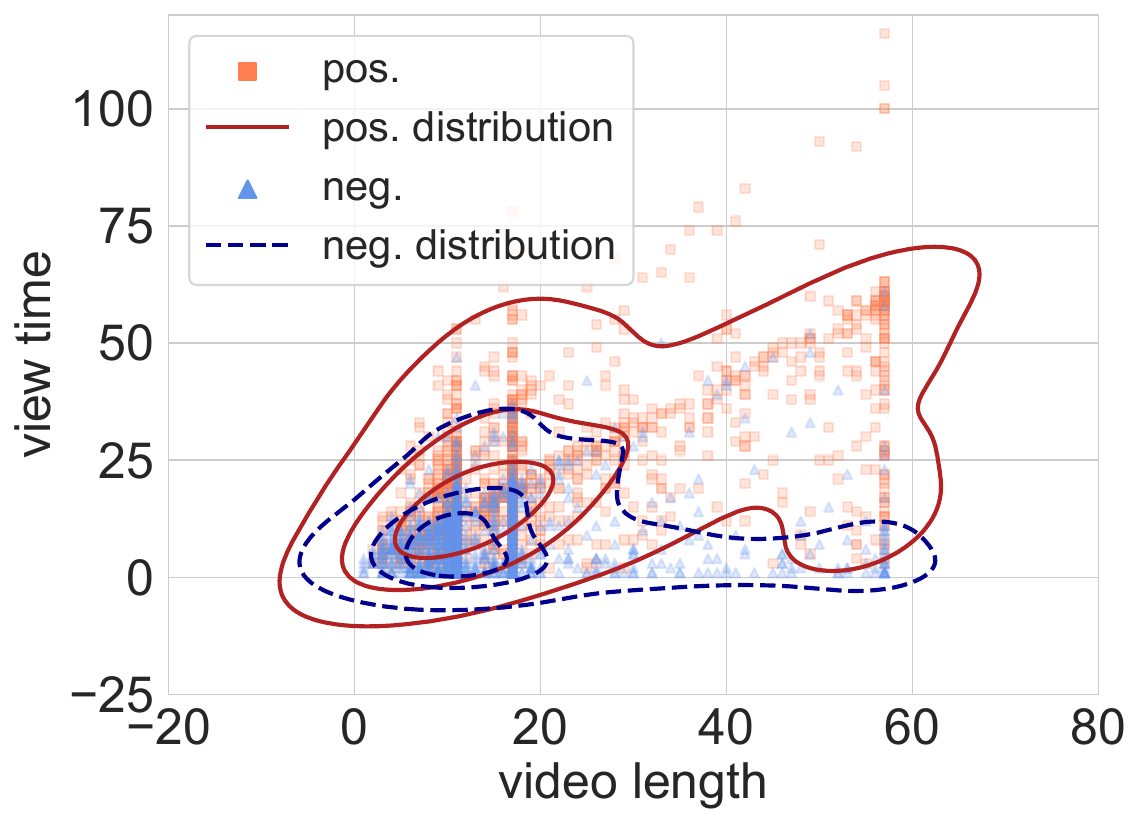}}
    \subfigure[Based on pointwise hard labeling]{\includegraphics[width=.49\textwidth]{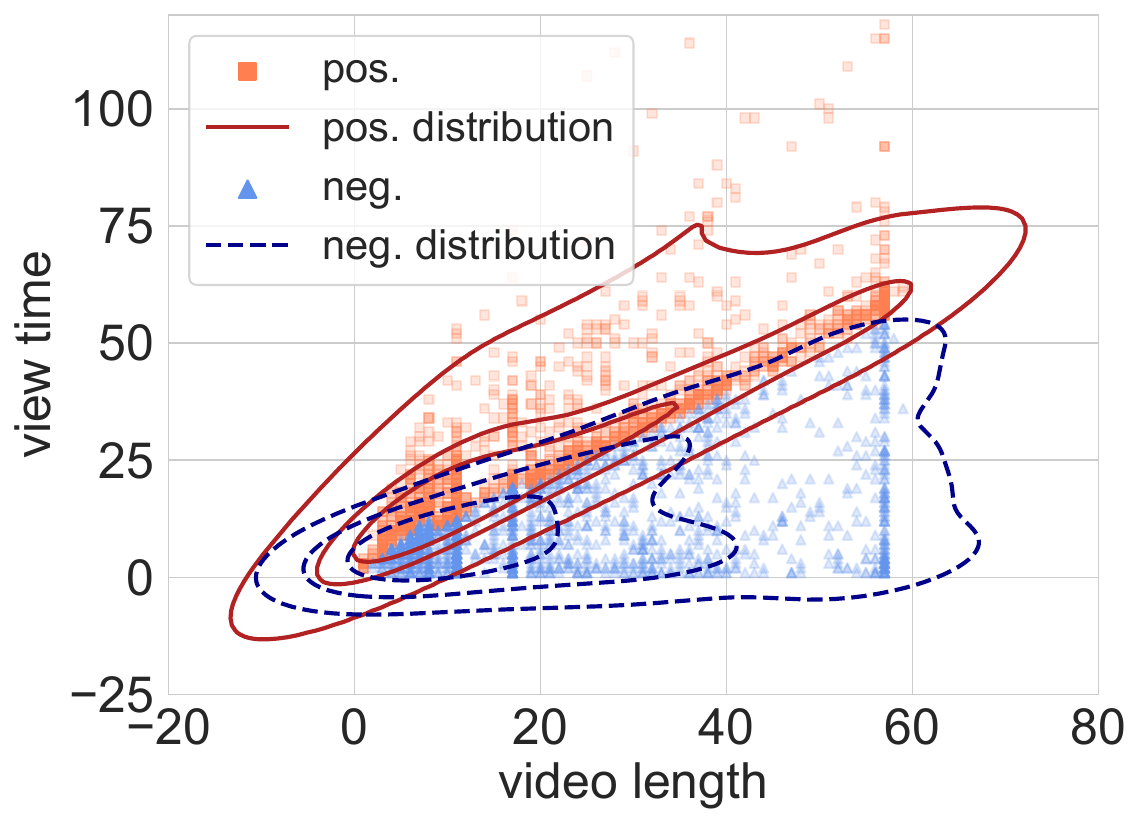}}
    \subfigure[Based on pairwise margin-based labeling]{\includegraphics[width=.49\textwidth]{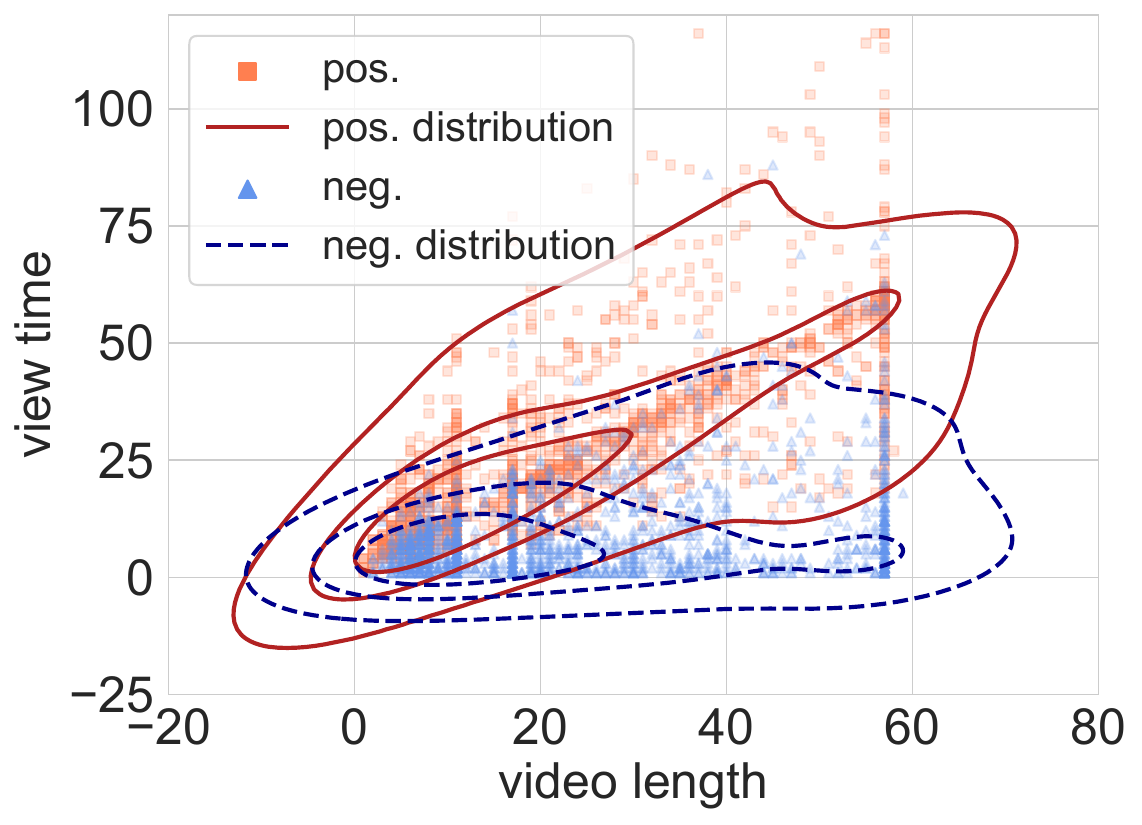}}
    \caption{Correlations between video length and view time of positive~(or negative) samples under different labeling strategies.}
    \label{fig:pair_generation}
\end{figure}

In Fig.~\ref{fig:pair_generation}, we plot the positive and negative sample distribution under the above three labeling strategies, respectively, in terms of both video length~(x-axis) and view time~(y-axis). Specifically, we uniformly sample 2,000 training samples in one epoch to obtain the population.
As illustrated in Fig.~\ref{fig:pair_generation}(a), negative samples tend to be located in the bottom-left corner, \emph{i.e.}, with both short video length and short view time.
While for our proposed two strategies, the distribution of negative samples extends from bottom left to upper right, indicating a better discriminative capability of the learned model, as videos with long duration and long view time can also be possibly chosen as negative samples. 
Compared with the pointwise hard labeling~(Fig.~\ref{fig:pair_generation}(b)), the pairwise margin-based labeling~(Fig.~\ref{fig:pair_generation}(c)) generates more similar distributions among positive samples and negative samples, which both have multiple centers of both positive and negative samples, and can serve as a complement to the former.

\subsection{Length-conditioned Sample Generation}
Our previous observation shows that completion rate distribution stays stable conditioned on the video length. Thus it is reasonable to draw a conclusion that, among two micro-videos with similar time length, a user favors the one with higher play progress~(\emph{i.e.}, longer view time).
Inspired by this, we are able to alleviate biased video length effect by learning pairwise rank relations among videos within the same group, which is achieved by a length-conditioned sample generation module.
Specifically, for each sample $(u_i,v_j)$, we additionally choose a video $v^{un}_k$ from the group corresponding to video $v_j$ to construct another training pair. The labels of $v_j$ and $v^{un}_k$ are defined similarly as introduced before.

The complete process of generating training samples is shown in Algorithm ~\ref{alg-sample}, which contains two parts of samples, one with video length effect and the other with alleviated effect.
Specifically, if current training pair uses the pointwise hard labeling, the candidate sets for sampling $S^{-}_{u_i,v_j}$ are $\{(u_i,v_k)|(t_{ik}<\tau(g) \land t_{ij}>\tau(g)) \lor (t_{ik}>\tau(g) \land t_{ij}<\tau(g))\}$ and $\{(u_i,v^{un}_k)|(t_{ik}<\tau(g) \land t_{ij}>\tau(g) \land v^{un}_k \in g(v_j)) \lor (t_{ik}>\tau(g) \land t_{ij}<\tau(g) \land v^{un}_k \in g(v_j)) \}$.
Otherwise, current training pair uses the pairwise margin-based labeling, and $S^{-}_{u_i,v_j}$ are $\{(u_i,v_k)| |p_{ik}-p_{ij}|>\epsilon \}$ and $\{(u_i,v^{un}_k)| |p_{ik}-p_{ij}|>\epsilon \land v^{un}_k \in g(v_j) \}$.

\begin{algorithm}[t]
	\caption{Length-conditioned Sample Generation Algorithm}
	\label{alg-sample}
	\begin{algorithmic}[1]
	\REQUIRE Positive instance $(u_i,v_j)$, where $v_j\in g_v$ and hyper parameter $\beta$, $\tau$, $\epsilon$
	\STATE p = rand()
	\IF{$p<\beta$} 
	\STATE //General sample
	\STATE sample $v_k$ from $S_{u_i}$ where $(t_{ik}<\tau(g) \land t_{ij}>\tau(g)) \lor (t_{ik}>\tau(g) \land t_{ij}<\tau(g))$
	\STATE // Length-conditioned
	\STATE sample $v^{un}_k$ from $S_{u_i} \cap g_v$ where $(t_{ik}<\tau(g_v) \land t_{ij}>\tau(g_v)) \lor (t_{ik}>\tau(g_v) \land t_{ij}<\tau(g_v))$
	\ELSE
	\STATE //General sample
	\STATE sample $v_k$ from $S_{u_i}$ where $\left |p_{ij}-p_{ik}\right| > \epsilon$
	\STATE // Length-conditioned
	\STATE sample $v^{un}_k$ from $S_{u_i} \cap g_v$ where $\left |p_{ij}-p_{ik}\right | > \epsilon$
	\ENDIF
    \RETURN  Positive instance $(u_i,v_j)$ and negative instances $(u_i,v_k)$, $(u_i,v_k^{un})$
    \end{algorithmic}
\end{algorithm}

\subsection{Multi-task User Preference Learning}
In this part, we design a multi-task learning model that enables joint optimization with two parts of samples.
Specifically, it contains a shared embedding module and two independent feedforward neural networks~(FNN) used for learning and predicting user preference from two types of samples, respectively.
First, for instances $(u_i,v_j)$, $(u_i, v_k)$,$(u_i, v_k^{un})$, the shared embedding module outputs $(\textbf{u}_i,\textbf{v}_j)$, $(\textbf{u}_i, \textbf{v}_k)$ and $(\textbf{u}_i, \textbf{v}^{un}_k)$ through a embedding map. These embeddings are further constructed into two training pairs, \emph{i.e.}, $\{(\textbf{u}_i,\textbf{v}_j), (\textbf{u}_i, \textbf{v}_k)\}$ and $\{(\textbf{u}_i,\textbf{v}_j), (\textbf{u}_i, \textbf{v}^{un}_k)\}$.
Then, the corresponding FNN for each training pair can generate the preference scores that are used for training or prediction. Note that these two networks do not share model parameters, and can be any commonly used recommendation models like NFM~\cite{he2017neural}, DeepFM~\cite{guo2017deepfm}, AutoInt~\cite{song2019autoint} etc.
Finally, a multi-task learning based training process is conducted to jointly learn user preference from both two parts of training samples, \emph{i.e.}, one with biased observation affected by video length and the other with debiased manipulation. 
Here we use the aforementioned BPR loss~\cite{rendle2012bpr}, and then add the two losses by linear weighting to get the final loss and use it for backpropagation. The formula is as follows,
\begin{equation}
\begin{aligned}
    & L_1 = L_{BPR}(f(\textbf{u}_i, \textbf{v}_j), f(\textbf{u}_i, \textbf{v}_k)),\quad  \\
    & L_2 = L_{BPR}(f_{un}(\textbf{u}_i, \textbf{v}_j), f_{un}(\textbf{u}_i, \textbf{v}^{un}_k)), \quad \\
    & L = \alpha*L_1 +(1-\alpha)*L_2,
\end{aligned}
\label{equ:loss_func}
\end{equation}
where $f$ and $f_{un}$ respectively denote the model used for two parts of training data. 

So far, we have completed the training process of VLDRec, and the overall procedure is shown in Algorithm~\ref{alg-overall}.

\begin{algorithm}[t]
	\caption{Overall procedure of VLDRec}
	\label{alg-overall}
	\begin{algorithmic}[1]
	\REQUIRE $U$, $V$, $\{l_j\}$ and training instances $\{u_i, v_j, t_{ij}\}$, Randomly initialize $\theta$
	\STATE calculate $Completion\_rate_v$ for $v \in V$
	\STATE generate video groups $\{g_v\}$
	\WHILE{$Stopping$  $criteria$ $is$ $not$ $met$}
	    \STATE generate negative instances $(u_i,v_k)$, $(u_i,v_k^{un})$ by Algorithm~\ref{alg-sample} for positive instance $(u_i, v_j)$
	    \STATE calculate $L_1$, $L_2$ and $L$ by Equation~\ref{equ:loss_func}
        \STATE update model parameter $\theta$ by minimizing $L$
	\ENDWHILE
    \end{algorithmic}
\end{algorithm}

\subsection{Length-invariant Top-$T$ Evaluation Metric}\label{sec: metric}

In both literatures and practices on personalized recommender systems~\cite{wang2019neural,yao2021self}, Top-$K$ based metrics are widely used for evaluating models, such as $Recall@K$ or $NDCG@K$.
However, in micro-video recommendation scenarios, as mentioned above, users can only watch one video at a time and the video is automatically played, so the objective is to maximize users' total view time, which depends on whether the recommendation model can capture the real user preference instead of the distorted one based on biased observations~(\emph{i.e.}, longer videos receive longer view time.).
Specifically, imagine an extreme case where a model always recommends micro-videos with a long length, a user's $View\_Time@K$, which is defined as
\begin{equation}
    View\_Time@K = \sum_{j=1}^{K}t_{ij}, 
\end{equation}

may still be fairly good even if she completes none of the recommended micro-videos. 
Therefore, we argue that Top-$K$ based metric is not suitable for view-time oriented recommendation tasks.

To better evaluate micro-video recommendation models from a fair angle, we propose a new Top-$T$ recommendation protocol where each time a list of videos are recommended and the total length of these videos is fixed as $T$, while the list length $K$ can be a variable.
Mathematically, the metric $View\_Time@T$ is defined as follows,
\begin{equation}
    \begin{aligned}
        & View\_Time@T = \sum_{j=1}^{n}t_{ij},  \\
        & where \sum_{j=1}^{n}l_j=T.
    \end{aligned}
\end{equation}
If the total video length exceeds $T$, the last video length and the corresponding user's view time will be adjusted proportionally, so that the total video length exactly equals $T$.
Compared with $View\_Time@K$, $View\_Time@T$ emphasizes the importance of play progress, which penalizes the extreme case where only long videos are recommended but the corresponding play progress is low.
However, it is noteworthy that, though similar, $View\_Time@T$ is intrinsically different from the average progress metric, as it gives a comprehensive account of both play progress and total view time.

\subsection{Discussion}\label{sec: discussion}
Regarding the video length bias problem in micro video recommendations, as mentioned above, there have been some related works. While the methods used may vary, the common objective is to mitigate the influence of video length on the recommendation model. From the perspective of causal inference, video length is a confounder between user and video. Therefore, the key to alleviating the bias is to alleviate the impact of the confounder. Existing related works attempt to achieve this by employing backdoor adjustment ~\cite{zhan2022deconfounding} or designing prediction targets independent of video length~\cite{zheng2022dvr}.

In our research, we conducted empirical data analysis and discovered that longer videos tend to have a higher average view time. Interestingly, this finding has also been directly utilized as prior information in two other related works~\cite{zhan2022deconfounding, zheng2022dvr}. Consequently, our approach to mitigating the confounding effect involves generating length-conditioned samples to construct unbiased data sets. Notably, this idea aligns with the theoretical support for causal inference, which is also shared by the related works.

\section{Experiments}
\label{sec: experiment}

\begin{table}[]
    \centering
    \caption{Basic information of datasets.}
    \begin{tabular}{c|cc}
        \toprule
         Dataset & Kuaishou & Wechat \\
         \midrule
         \#Samples & 1,945,502 & 3,264,803 \\
         \#Users & 9,829 & 54,595 \\
         \#Videos & 136,317 & 62,569 \\
         Average video length & 17.54s & 32.97s \\
         Max video length & 60s & 120s \\
         Average view time & 14.78s & 26.32s \\
         \bottomrule
    \end{tabular}
    \label{Dataset}
\end{table}

We aim to answer the following three research questions (RQ) in experiments.
\begin{itemize}[leftmargin=*]
    \item \textbf{RQ1}: How does our proposed VLDRec model perform compared with state-of-the-art micro-video recommendation methods. More specifically, does VLDRec successfully alleviate the video length effect that troubles common practice in previous solutions?
    \item \textbf{RQ2}: Whether the modules of our model can work well, including the debias sampling strategy and the multi-task learning based model design.
    \item \textbf{RQ3}: Can the VLDRec capture user preference on micro-videos, in terms of other dimensions besides view time?
\end{itemize}

\subsection{Experimental Settings}

\subsubsection{Dataset and data preprocessing}
We conduct extensive experiments on two real-world datasets collected from popular micro-video applications, Kuaishou and Wechat Channels. 
The Kuaishou dataset is a public dataset\footnote{https://github.com/liyongqi67/ALPINE} that has been used in the previous work~\cite{li2019routing}. Since the optimization goal of this work is the user’s click behavior rather than the viewing time, which is different from our task, we do not use this work as our baseline. Wechat dataset is collected from real industrial scenarios. Besides, we only keep the instances that users have clicked and watched, as those non-click videos are not watched and do not match the optimization objective in the micro-video recommendation task. It should be noted that in the original data, the WeChat dataset contains timestamps but the Kuaishou dataset does not contain them. Therefore, data analysis like Table~\ref{fig:comparision_records} can only be performed on the WeChat dataset.

It is noteworthy that in micro-video applications, each video is automatically played. Unless the user slides down to the next video, the current video will be played repeatedly. Therefore, samples that are played too many times may be abnormal and we deleted samples where the video is played more than 3 times($p_{ij}>3$) repeatedly in two datasets. At the same time, according to the difference among the datasets, we deleted a small number of samples where the video length is very long. In the two datasets, the threshold is set to 60 seconds and 120 seconds, respectively. 

Table~\ref{Dataset} summarizes some basic information about the datasets we used. Since the two datasets do not contain the timestamp, we randomly split 10\% as the validation set and 20\% as the test set. According to the length of the video, we divided the Kuaishou and Wechat datasets into 5 and 7 groups respectively. More details about grouping have been described in section~\ref{sec:ltr-approach}. The features we use include $user\_id$, $video\_id$ and $video\_length$, as we aim to evaluate the proposed VLDRec model in a general experimental setting.

\subsubsection{Baseline}
We compare the proposed VLDRec with three categories of baseline methods.

The first category is two regression-based methods that are widely used in industrial micro-video recommender systems.
\begin{itemize}[leftmargin=*]
    \item \textbf{TimeRegression~(TReg)}. This is an intuitive method that takes the length of time the user watched the video as the target($t_{ij}$), and sorts according to the predicted view time, so as to obtain the final recommendation result. 
    \item \textbf{RateRegression~(RReg)}. This baseline regresses the view progress of each sample($p_{ij}$), and then the predicted view time is obtained by multiplying the predicted score by the video length.
\end{itemize}

The next category is two ranking-based methods that follow a similar idea of learning-to-rank as ours in solving micro-video recommendation task.
\begin{itemize}[leftmargin=*]
    \item \textbf{TimeRanking~(TRank)}. This method learns to rank according to the view time of the instances, where the negative instances are randomly sampled from the videos watched by the user of the positive instance. Based on our proposed multi-task learning framework, only biased datasets are used for ranking training in this method, that is, the weight of multi-task is set to 0. It is a de-generated model that there is no unbiased dataset to help learn unbiased representations, directly ranking the samples on the biased dataset may cause a large bias in the recommended model obtained by training.
    
    \item \textbf{RateRanking~(RRank)}. This method learns to rank according to the play progress of the instances and its sampling strategy is consistent with the \textit{TimeRanking} method. It should be noted that since the prediction score obtained by the ranking method is not an accurate value, it does not need to be multiplied by the video length, but is directly used for ranking and recommendation.
\end{itemize}

The final category is five unbiased recommendation methods that try to learn an unbiased recommender model by removing the video length effect.
\begin{itemize}[leftmargin=*]
    \item \textbf{IPS}~\cite{joachims2017unbiased, schnabel2016recommendations}. The inverse propensity score method estimates the bias by re-weight each instance. For those items that have a greater advantage in recommendation due to bias, the IPS method reduces the weight of these instances to achieve a balance in recommendation results. In our experiment, the weight of each instance is set to the inverse of the video length of each instance; that is, the longer the video length, the lower the weight of the instance. 
    \item \textbf{IPS-C}~\cite{bottou2013counterfactual}. This method uses max capping to limit the value of IPS, so that the range of IPS values is limited, thereby reducing the variance of the score and enhancing the stability of the model
    \item \textbf{IPS-CN}~\cite{gruson2019offline}. On the basis of max capping, this method normalizes the value of IPS, so that the variance of the IPS score is further reduced.
    \item \textbf{IPS-CNSR}~\cite{gruson2019offline}. On the basis of normalization, this method adds smoothing operations for the value of IPS.
    \item \textbf{CausE}~\cite{CausE}. It is trained through a large biased dataset and a small unbiased dataset. By using two models to model two datasets respectively, and using L2 regularization to constrain the embedding of the two models, so the impact of the bias of the model can be reduced. In our experiment, the unbiased dataset is obtained by sampling from the same group of the video of the positive instance, which is consistent with part of our VLDRec model.
    \item \textbf{DecRS}~\cite{wang2021deconfounded}. The algorithm avoids the impact of confounder by inserting a backdoor adjustment operator into the existing model and solves the problem of infinite sample space through an approximation algorithm.
    \item \textbf{DVR}~\cite{zheng2022dvr}. This algorithm designs a new unbiased metric named WTG as the prediction target, and trains the model through adversarial learning.

\end{itemize}

\subsubsection{Evaluation Method}

Since the proposed VLDRec is mainly aimed at the ranking stage in recommendation, we only use the collected records to construct recommendation lists.
For the samples in the test set, we first aggregated them according to $user\_id$. Then, for each user, a recommendation list is generated by sorting candidate micro-videos according to prediction scores and selecting the top ones. The final evaluation metrics are obtained by first calculating per-user value and then averaging among all users. 

First of all, we mainly use view time metrics in experiments, including $View\_Time@T$ and $View\_Time@K$, where the former is less affected by biased observations and formally defined in Section~\ref{sec: metric}. 
Although $View\_Time@K$ favors longer videos and cannot reflect users' real preferences, we still use this metric to highlight the spurious goodness-of-fit of prevailing solutions.

Moreover, besides the view time, we also leverage category information of micro-videos to measure whether our recommendation reflects users' actual preferences.
Specifically, for Top-$K$ recommended videos of each user, we calculate \textit{size of intersection} and Jensen–Shannon divergence~($JSD$)~\cite{fuglede2004jensen} by comparing them with the actual Top-$K$ viewed videos ordered by view time.
The first metric \textit{size of intersection} is in $[0,K]$. 
As for $JSD$, it is defined as 
\begin{equation*}
    \text{JSD}(P||Q) = H(\frac{P+Q}{2})-\frac{1}{2}(H(P)+H(Q))
\end{equation*}
where $H$ is the Shannon entropy, $P$ and $Q$ are distributions. Lower $JSD$ denotes a closer distribution between the recommendation and real data, indicating a better reflection of user preference.

\subsubsection{Implementation Detail}

For all methods, we use NFM~\cite{he2017neural}, DeepFM~\cite{guo2017deepfm} and AutoInt~\cite{song2019autoint} as base model. The embedding size for each feature of all methods is set to 8 and the batch size is set to 1024. The number of hidden layers of the deep part of NFM model is set to $\{32,16\}$. For the regression method, we use MSE loss. In all experiments, we use the Adam optimizer. Hyper parameters such as learning rate and dropout are obtained through grid search. The search ranges are shown in Table~\ref{tab:hp_range}.
The code and dataset are avaliable\footnote{https://github.com/SpongeBobSquarePants111/VLDRec}.

\begin{table*}[]
    \setlength\tabcolsep{5pt}
    \centering
    \caption{Search range of hyper-paramaters.}
    \begin{tabular}{c|c}
        \toprule
         Hyper-parameter  & Range \\
         \midrule
          $lr$ & $\{0.005,0.001,0.0005,0.0001\}$ \\
          $dropout$ & $[0,1]$ \\
          $\alpha$ & $\{0.1,0.3,0.5,0.7,0.9\}$ \\
          $\beta$ & $\{0.1,0.3,0.5,0.7,0.9\}$ \\
         \bottomrule
    \end{tabular}
    \label{tab:hp_range}
\end{table*}

\subsection{Performance Comparison~(RQ1)}

\subsubsection{Overall Performance}

\begin{table*}[]
    \setlength\tabcolsep{5pt}
    \centering
    \caption{Overall performance comparison between our proposed method and baselines on Kuaishou with NFM as base model(RelaImpr is shorthand for Relative Improvement).}
    \begin{tabular}{c|cccc}
        \toprule
         Metric  & View\_time@120 & RelaImpr & View\_time@240 & RelaImpr \\
         \midrule
         TReg & 35.78 & 25.66\% & 66.69 & 23.51\%   \\
         RReg & 25.13 & 78.91\% & 37.32 & 120.71\%  \\
         TRank    & 40.87 & 10.01\% & 78.79 & 4.54\%   \\
         RRank    & 43.94 & 2.32\%  & \underline{81.93} & 0.54\% \\
         IPS            & 40.31 & 11.54\% & 77.51 & 6.27\%  \\
         IPS-C          & 43.36 & 3.69\%  & 80.99 & 1.70\%  \\
         IPS-CN         & 41.62 & 8.02\%  & 79.03 & 4.23\%   \\
         IPS-CNSR       & 41.80 & 7.56\%  & 79.75 & 3.29\%   \\
         CausE          & 42.14 & 6.69\%  & 80.11 & 2.82\%   \\
         DecRS          & \underline{44.16} & 1.81\% & 71.69 & 14.90\% \\
         DVR        & 40.22 & 10.54\% & 79.47 & 3.52\% \\
         \midrule
         VLDRec     & \textbf{44.96} & - & \textbf{82.37} & -\\
         \bottomrule
    \end{tabular}
    \label{tab:overall_performance_kuaishou_nfm}
\end{table*}

\begin{table*}[]
    \setlength\tabcolsep{5pt}
    \centering
    \caption{Overall performance comparison between our proposed method and baselines on Kuaishou with DeepFM as base model(RelaImpr is shorthand for Relative Improvement).}
    \begin{tabular}{c|cccc}
        \toprule
         Metric  & View\_time@120 & RelaImpr & View\_time@240 & RelaImpr \\
         \midrule
         TReg & 25.01 & 92.04\% & 52.13 & 67.10\%   \\
         RReg & 25.75 & 86.52\% & 41.37 & 110.56\%  \\
         TRank    & \underline{45.53} & 5.49\% & \underline{83.39} & 4.46\%   \\
         RRank    & 38.27 & 25.50\%  & 79.31 & 9.83\% \\
         IPS            & 35.52 & 35.22\% & 75.70 & 15.07\%  \\
         IPS-C          & 36.01 & 33.28\%  & 76.50 & 13.87\%  \\
         IPS-CN         & 35.58 & 34.99\%  & 76.50 & 13.87\%   \\
         IPS-CNSR       & 37.61 & 27.71\%  & 79.06 & 10.18\%   \\
         CausE          & 36.19 & 32.72\%  & 76.84 & 13.37\%   \\
         DecRS          & 37.26 & 28.90\% & 63.05 & 38.16\% \\
         DVR        & 38.01 & 20.86\% & 83.38 & 4.28\% \\
         \midrule
         VLDRec     & \textbf{48.03} & - & \textbf{87.11} & -\\
         \bottomrule
    \end{tabular}
    \label{tab:overall_performance_kuaishou_deepfm}
\end{table*}

\begin{table*}[]
    \setlength\tabcolsep{5pt}
    \centering
    \caption{Overall performance comparison between our proposed method and baselines on Kuaishou with AutoInt as base model(RelaImpr is shorthand for Relative Improvement).}
    \begin{tabular}{c|cccc}
        \toprule
         Metric  & View\_time@120 & RelaImpr & View\_time@240 & RelaImpr \\
         \midrule
         TReg & 40.09 & 17.56\% & 65.68 & 21.92\%   \\
         RReg & 22.36 & 110.78\% & 35.00 & 128.80\%  \\
         TRank    & 37.37 & 26.12\% & 66.16 & 21.04\%   \\
         RRank    & 34.23 & 37.69\%  & 61.72 & 29.75\% \\
         IPS            & 41.86 & 12.59\% & 71.75 & 11.61\%  \\
         IPS-C          & 40.39 & 16.69\%  & 69.38 & 15.42\%  \\
         IPS-CN         & 40.38 & 16.72\%  & 69.53 & 15.17\%   \\
         IPS-CNSR       & \underline{43.92} & 7.31\%  & 74.90 & 6.92\%   \\
         CausE          & 42.55 & 10.76\%  & 72.20 & 10.91\%   \\
         DecRS          & 43.18 & 9.15\% & \underline{76.02} & 5.34\% \\
         DVR        & 38.15 & 19.05\% & 74.21 & 7.33\% \\
         \midrule
         VLDRec     & \textbf{47.13} & - & \textbf{80.08} & -\\
         \bottomrule
    \end{tabular}
    \label{tab:overall_performance_kuaishou_autoint}
\end{table*}

\begin{table*}[]
    \setlength\tabcolsep{5pt}
    \centering
    \caption{Overall performance comparison between our proposed method and baselines on Wechat with NFM as base model(RelaImpr is shorthand for Relative Improvement).}
    \begin{tabular}{c|cccc}
        \toprule
        Metric  & View\_time@120 & RelaImpr & View\_time@240 & RelaImpr \\
         \midrule
         TReg & 12.52 & 137.30\% & 27.16 & 97.05\% \\
         RReg & 12.17 & 144.12\% & 21.49 & 149.05\% \\
         TRank & 23.50 & 26.43\%  & 48.10 & 11.27\% \\
         RRank & 22.84 &      30.08\%  & 47.01 &  13.85\% \\
         IPS    & 22.73 & 30.71\%  & 46.26 & 15.69\% \\
         IPS-C  & 22.45 & 32.34\%  & 47.65 & 12.32\% \\
         IPS-CN & 23.53 & 26.26\%  & 47.40 & 12.91\% \\
         IPS-CNSR   & 23.55 & 26.16\%  & 47.07 & 13.70\% \\
         CausE      & \underline{26.69} & 11.32\%  & \underline{51.59} & 3.74\% \\
         DecRS      & 24.54 & 21.06\% & 43.23 & 23.80\% \\
         \midrule
         VLDRec & \textbf{29.71} & - & \textbf{53.52} & -\\

         \bottomrule
    \end{tabular}
    \label{tab:overall_performance_wechat}
\end{table*}

The overall experimental results are shown in the Table~\ref{tab:overall_performance_kuaishou_nfm}, Table~\ref{tab:overall_performance_kuaishou_deepfm}, Table~\ref{tab:overall_performance_kuaishou_autoint} and Table~\ref{tab:overall_performance_wechat} \emph{w.r.t.} $View\_Time@T$. From the above results, we have the following observations:
\begin{itemize}[leftmargin=*]
    \item \textbf{VLDRec significantly improves the recommendation performance by alleviating the video length effect and capturing real user preferences regardless of the video length.} Compared with the best-performed baseline, it outperforms by 1.81\% and 11.32\% \emph{w.r.t.} $View\_Time@120$ in Kuaishou and Wechat with NFM as the base model, respectively. At the same time, when using DeepFM and AutoInt as the base model, VLDRec also achieved at least 5.49\% and 7.31\% improvements \emph{w.r.t.} $View\_Time@120$, respectively. It demonstrates that actively removing the biased effects introduced by video length is vital for learning user preferences among different micro-videos.
    
    \item \textbf{VLDRec better resists the noisy preference signal in biased observation.} Corresponding to our aforementioned problem of using view time based signal, we observe the significant performance degradation with two regression methods that are common practice in most companies, \emph{i.e.}, TimeRegression and RateRegression. Specifically, VLDRec outperforms the best of them by 25.66\% in Kuaishou and 137.30\% in Wechat \emph{w.r.t.} $View\_Time@120$ with NFM as the base model. At the same time, VLDRec outperforms the best of them 92.04\% and 17.56\% with DeepFM and AutoInt as the base model. The observed huge performance gap demonstrates that directly using observed view time~(or play progress) as the label can be biased and lead to distorted preference learning. Contrastively, VLDRec successfully bypasses this by learning pairwise ranking relations of user preferences among micro-videos with similar time lengths.
    
    \item \textbf{VLDRec further improves the informativeness of training samples by selecting training pairs conditioned on video length.}
    By choosing pairs of micro-videos with similar time lengths, VLDRec not only alleviates the existing bias in observed data but also leverages the advantage of hard negative sampling that is proven to be useful in improving recommendation performance~\cite{zhang2013optimizing,ding2019reinforced}. Intuitively, for a specific sample $(u_i,v_j)$, choosing a negative sample from the same group as $v_j$ is much more informative for model learning than a different group where video length is shorter.
    Therefore, compared with those unbiased learning baselines including IPS-based, CausE, DecRS and DVR, VLDRec outperforms by a large margin. Specifically, IPS-based methods suffer from the large variance issue, resulting in unstable performance as shown in Table~\ref{tab:overall_performance_kuaishou_nfm}, Table~\ref{tab:overall_performance_kuaishou_deepfm}, Table~\ref{tab:overall_performance_kuaishou_autoint} and Table~\ref{tab:overall_performance_wechat}. As for CausE, DecRS, and DVR they perform fairly competitively but cannot beat VLDRec, as it is not elaborately designed for a view time oriented recommendation task. 

\end{itemize}

\subsubsection{Biased Learning of Common Practice.}

The above performance comparison has demonstrated the superiority of VLDRec in recommending micro-videos that indeed match user interest and thus generate user engagement.
In this part, we further answer the question on how VLDRec alleviates the video length effect that existed in previous solutions.

We first present the recommendation performances of each video-length group~(\emph{i.e.}, $\{g_v\}$) \emph{w.r.t.} $View\_Time@K$.
Specifically, for each user, only top-$K$ micro videos belonging to a certain group are recommended and $View\_Time@K$ is used for evaluation. As the recommended videos are of a similar time length, $View\_Time@K$ is freed from the aforementioned problem of favoring longer videos.
As shown in Table~\ref{tab:perf_group_kuaishou} and Table~\ref{tab:perf_group_wechat}, surprisingly, we observe that TimeRegression, RateRegression and DecRS perform rather competitively in each video group. Specifically, for Kuaishou Dataset, TimeRegression and RateRegression methods beat VLDRec in all five groups and DecRS beats VLDRec in four groups. As for Wechat Dataset, DecRS beats VLDRec in all groups, while TimeRegression outperforms in longer video groups~(60 - 120s) and RateRegression outperforms in shorter ones~(0 - 59s).
The above observation implies that these methods are able to learn user preferences among micro-videos with similar time lengths, even better than the proposed VLDRec. However, this is in conflict with the overall performance comparison in Table~\ref{tab:overall_performance_kuaishou_nfm}, Table~\ref{tab:overall_performance_kuaishou_deepfm}, Table~\ref{tab:overall_performance_kuaishou_autoint} and Table~\ref{tab:overall_performance_wechat}, where regression based methods are reported to suffer from significant performance degradation and DecRS doesn't perform well.

\begin{figure*}
\centering
    \subfigure[Mean normalized scores]{\includegraphics[width=.45\textwidth]{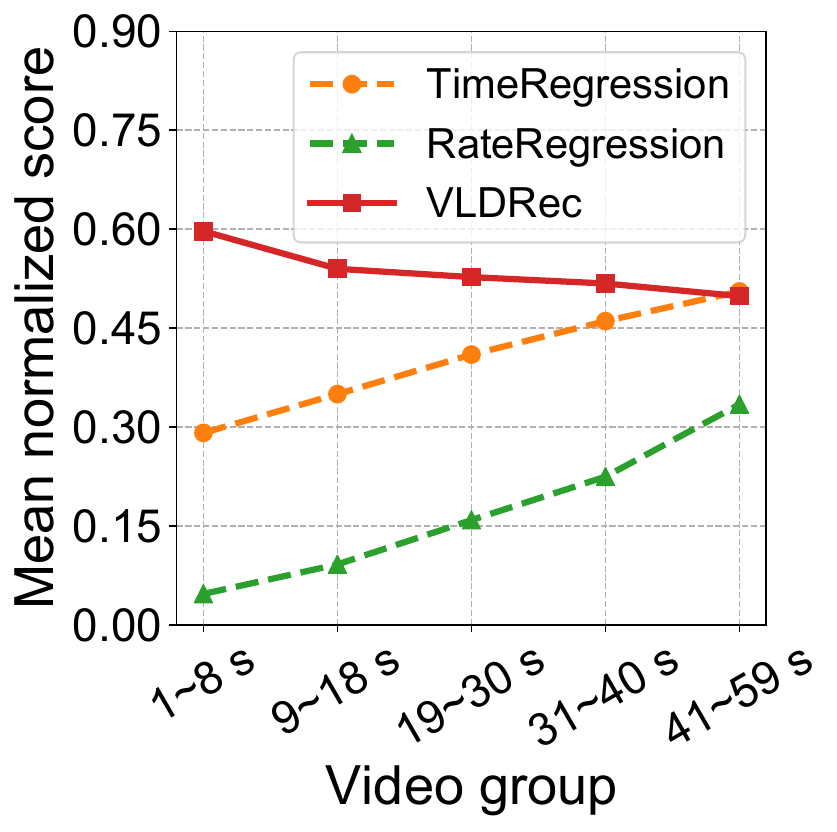}}
    \subfigure[Standard deviation]{\includegraphics[width=.45\textwidth]{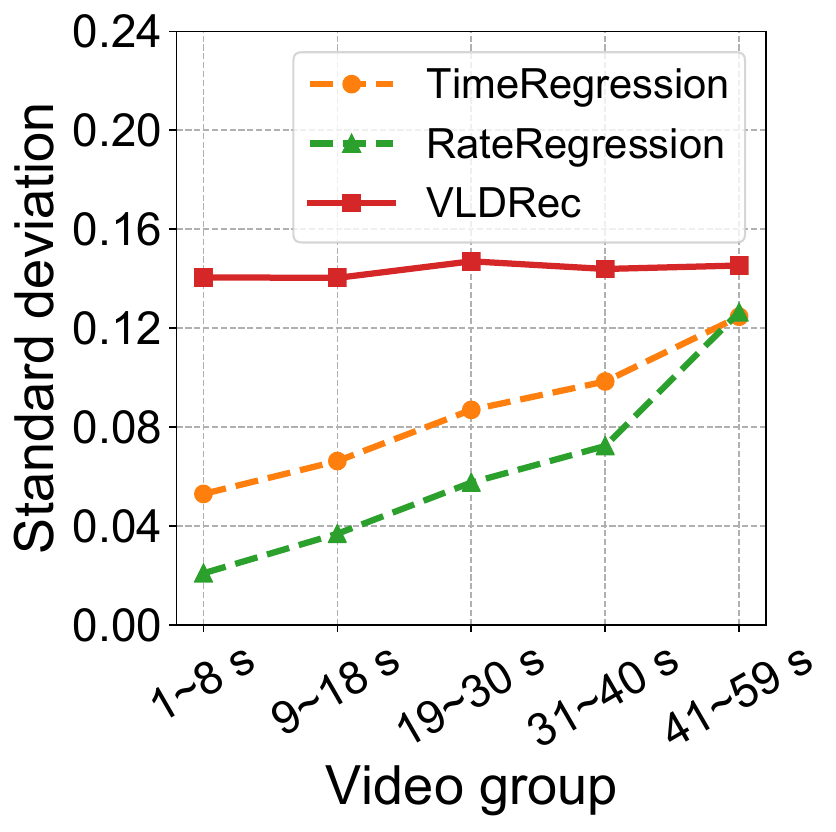}}
    \caption{Model prediction scores (mean normalized scores and standard deviation) under different video groups (divided by video length) on Kuaishou.}
    \label{fig:average_std_score_group_kuaishou}
\end{figure*}

\begin{figure*}
\centering
    \subfigure[Mean normalized scores]{\includegraphics[width=.45\textwidth]{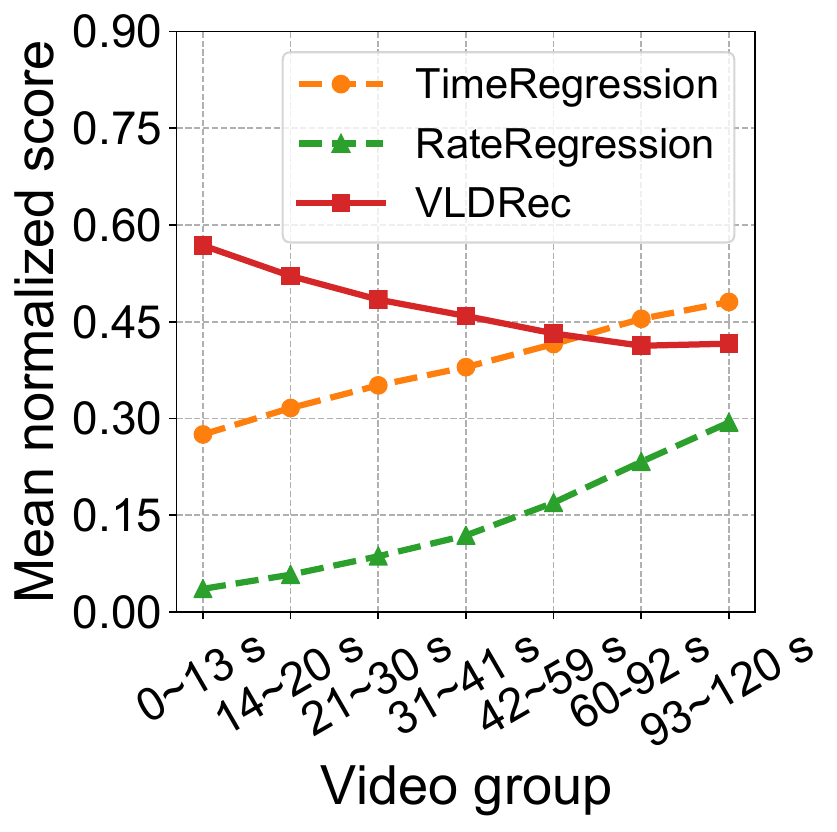}}
    \subfigure[Standard deviation]{\includegraphics[width=.45\textwidth]{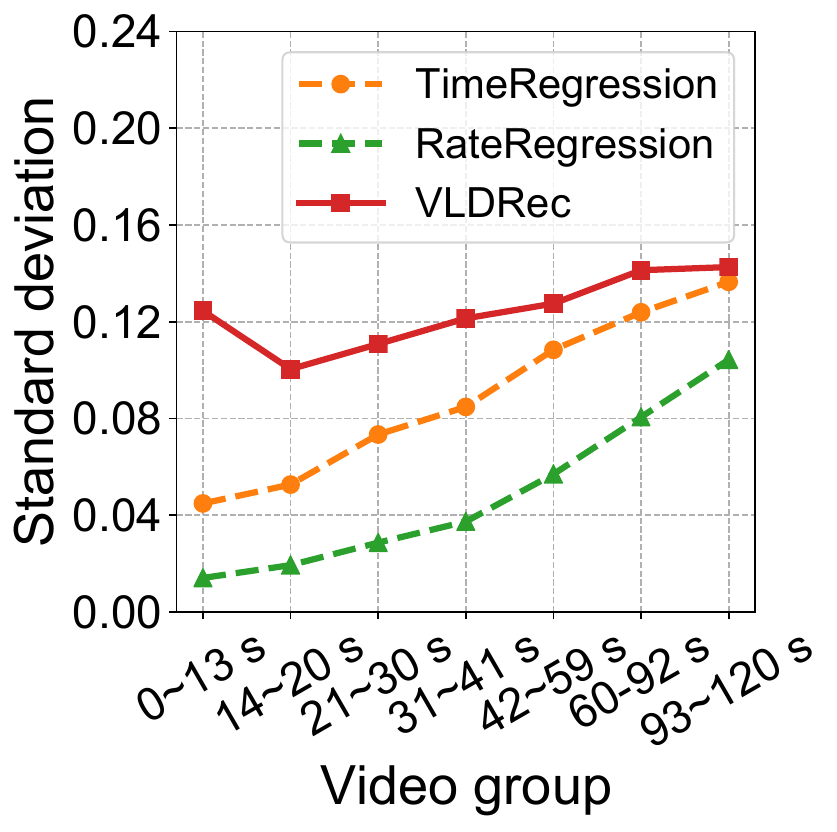}}
    \caption{Model prediction scores (mean normalized scores and standard deviation) under different video groups (divided by video length) on Wechat.}
    \label{fig:average_std_score_group_wechat}
\end{figure*}

To further explain the above contradictory phenomena, we analyze the distribution of model prediction scores generated by regression methods and VLDRec in Fig.~\ref{fig:average_std_score_group_kuaishou} and Fig.~\ref{fig:average_std_score_group_wechat}. Specifically, we normalize the prediction scores of all test set samples generated by each method and then calculate the mean score and standard deviation of each video group.
As shown in Fig. ~\ref{fig:average_std_score_group_kuaishou}, we can observe that the average prediction score of TimeRegression and RateRegression increases with the video time length, which is similar to the observation from empirical data that longer videos generally have longer watch time~(See Fig.~\ref{fig:average_view_time}.). This indicates a good fit to the training data, which, however, does not guarantee a precise recommendation that matches user preference. 
Further in Fig. ~\ref{fig:average_std_score_group_wechat}, when looking at the variation of scores~(\emph{i.e.}, standard deviation), the regression methods still have an increasing standard deviation as the video length increases, and low value of variation in short video groups indicates a centralized prediction distribution.
In summary, regression models achieve a good fit of distribution in terms of mean score and their predictions exhibit a rather centralized characteristic in terms of standard deviation, which means they are fairly precise in each video group but cannot overcome the intrinsic problem that longer videos tend to be ranked higher regardless of the actual user preference.

Contrastively, the proposed VLDRec has a relatively stable distribution of both mean score and standard deviation among all video groups.
Specifically, VLDRec generates almost equal mean scores among all video groups. Also, its standard deviation is higher than regression models, indicating that a micro-video assumed to be preferred by a certain user can be ranked higher even if this video just lasts less than 10s.
Therefore, VLDRec is less influenced by the video length and is able to distinguish micro-videos that users are truly interested in and not interested in. Comparatively, industrial recommender systems normally rely on ranking with a fusion of multiple objectives~(like a linear combination of two scores from TimeRegression and RateRegression) to alleviate the video-length effect in their recommendation results~\cite{mmoe}, which is less elegant and requires a huge manual effort for tuning fusion-related hyperparameters.

\begin{table}
    \centering
    \caption{Performance comparison \emph{w.r.t.} $View\_Time@K$~($K=3$) in different video groups on Kuaishou.}
    \begin{tabular}{c|ccccc}
    \toprule
    Group & 1$\sim$8 & 9$\sim$18 & 19$\sim$30 & 31$\sim$40 & 41$\sim$59 \\
    \midrule
    TReg & \underline{13.83} & 23.81 & \underline{44.68} & \underline{67.66} & \underline{73.31}  \\
    RReg & \textbf{22.95} & \textbf{35.30} & 38.26 & 51.41 & 47.71  \\
    TRank & 8.56 & 10.09 & 31.13 & 57.28 & 51.79 \\
    RRank & 9.07 & 10.91 & 32.09 & 56.67 & 50.94 \\
    IPS & 8.33 & 9.73 & 30.56 & 56.70 & 51.68 \\
    CausE & 8.67 & 10.38 &  31.52 & 56.99 & 51.75 \\
    DecRS & 9.51 & \underline{31.73} & \textbf{47.28} & \textbf{69.95} & \textbf{84.15}  \\
    DVR   & 14.08 & 26.28 & 43.20 & 65.71 & 72.62 \\
    \midrule
    VLDRec & 10.29 & 14.78 & 35.85 & 58.16 & 54.41  \\
    \bottomrule
    \end{tabular}
    \label{tab:perf_group_kuaishou}
\end{table}

\begin{table}
    \centering
    \caption{Performance comparison \emph{w.r.t.} $View\_Time@K$~($K=3$) in different video groups on Wechat.}
    \begin{tabular}{c|ccccccc}
    \toprule
    Group & 0$\sim$13 & 14$\sim$20 & 21$\sim$30 & 31$\sim$41 &42$\sim$59 & 60$\sim$92 & 93$\sim$120 \\
    \midrule
    TReg & 6.60 & 15.90& 28.45 & 45.31 & 49.98 & \underline{88.62} & \underline{125.17}  \\
    RReg & \textbf{29.57} & \textbf{37.07} & \textbf{45.60} & \underline{49.76} & \underline{59.34} & 64.81 & 70.20 \\
    TRank & 9.46 & 18.34 & 30.74 & 42.97 & 40.24 & 68.38 & 111.88\\
    RRank & 9.28 & 17.52 & 30.64 & 42.09 & 42.29 & 69.17 & 112.52\\
    IPS & 9.68 & 17.75 & 28.45 & 37.48 & 34.60 & 61.74 & 106.93\\
    CausE & 12.35 & 22.72 & 33.84 & 45.00 & 43.26 & 66.72 & 107.96\\
    DecRS & \underline{18.63} & \underline{27.48} & \underline{42.34} & \textbf{55.77} & \textbf{74.75} & \textbf{107.21} & \textbf{139.93} \\
    \midrule
    VLDRec & 15.82 & 26.14 & 35.98 & 46.89 & 49.23 & 72.66 & 115.67 \\
    \bottomrule
    \end{tabular}
    \label{tab:perf_group_wechat}
\end{table}

\subsection{Ablation Study~(RQ2)}

\subsubsection{Performance on Different Sampling Strategy}

In the process of instance generation, we proposed two sampling methods. The pointwise hard labeling based sampling and the pairwise margin-based labeling based sampling. The former focuses on characterizing the overall distribution between positive instances and negative instances, while the latter strengthens the modeling of preference rankings of the same user. We use hyper-parameter $\beta$ to control the sampling strategy. In order to verify the effect of the sampling strategy, we adjust the value of $\beta$ to observe changes in performance and show the result \emph{w.r.t.} $View\_Time@T$ in Fig.~\ref{fig:alpha_category_study}~(a). We can observe that the overall trend is rising first and then falling, which proves that the two sampling methods we proposed based on different objectives can effectively model the user's preferences and improve the performance of the model.

\subsubsection{Performance on Different Weight of Multitask Learning}
\begin{figure}[t!]
    \centering
    \subfigure[sampling strategy]{\includegraphics[width=.48\textwidth]{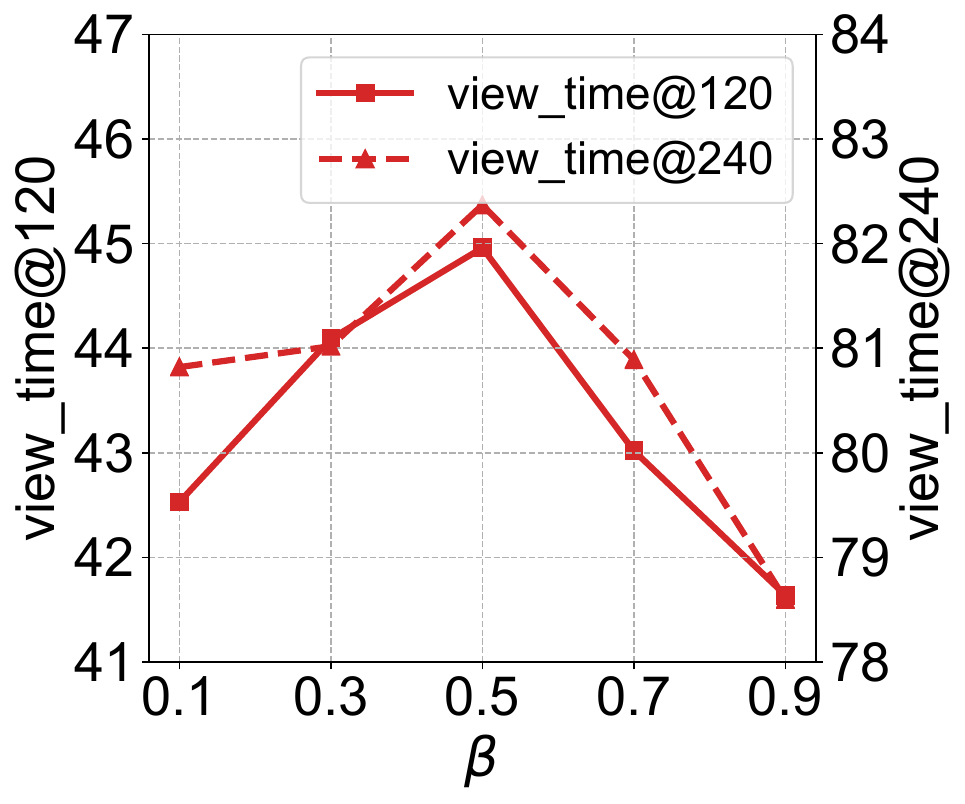}}
    \subfigure[multitask learning]{\includegraphics[width=.48\textwidth]{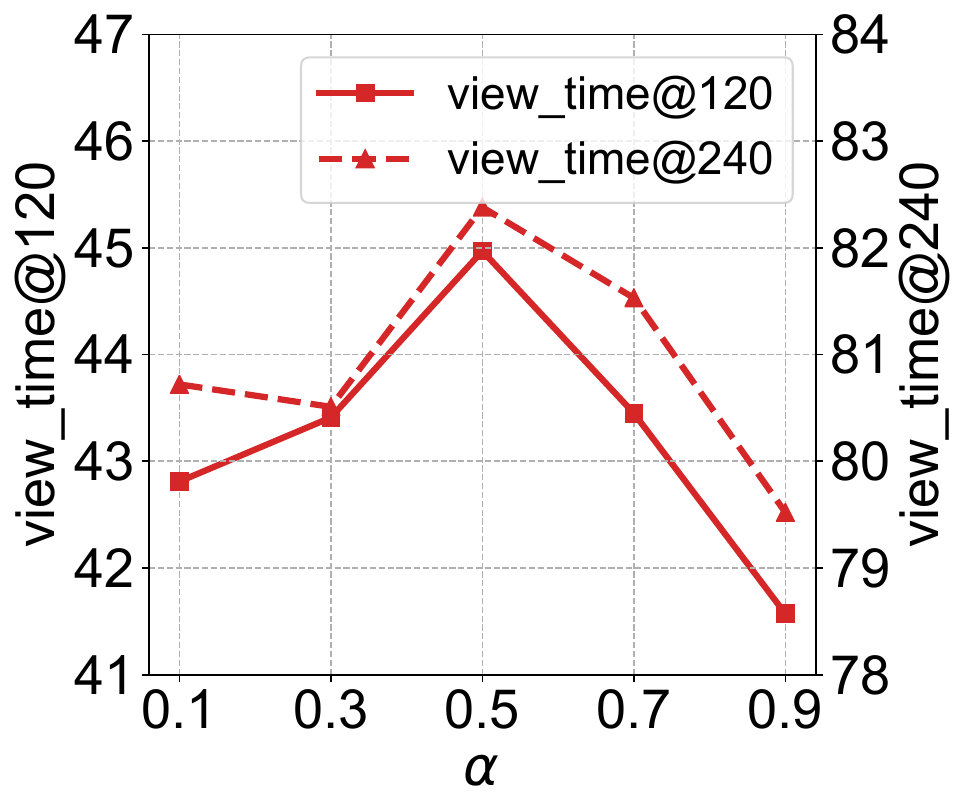}}
    \caption{Performance with different weight of sampling strategy and multitask learning in Kuaishou.}
    \label{fig:alpha_category_study}
\end{figure}

To alleviate the biased video length effect, VLDRec additionally learns to rank among positive and negative instances within the same video group and uses the multi-task learning strategy for training. To verify the effectiveness of this module, we adjust the hyper-parameter $\alpha$ of the multi-task learning strategy and illustrate the results in figure~\ref{fig:alpha_category_study}~(b). It can be observed that the performance of the model is the best when $\alpha=0.5$. Model performance degrades when $\alpha$ increases or decreases, which means that the model has achieved a balance between the two goals of alleviating bias and fitting the data distribution, thereby improving the recommendation performance.

\begin{figure}[t!]
    \centering
    \includegraphics[width=.7\textwidth]{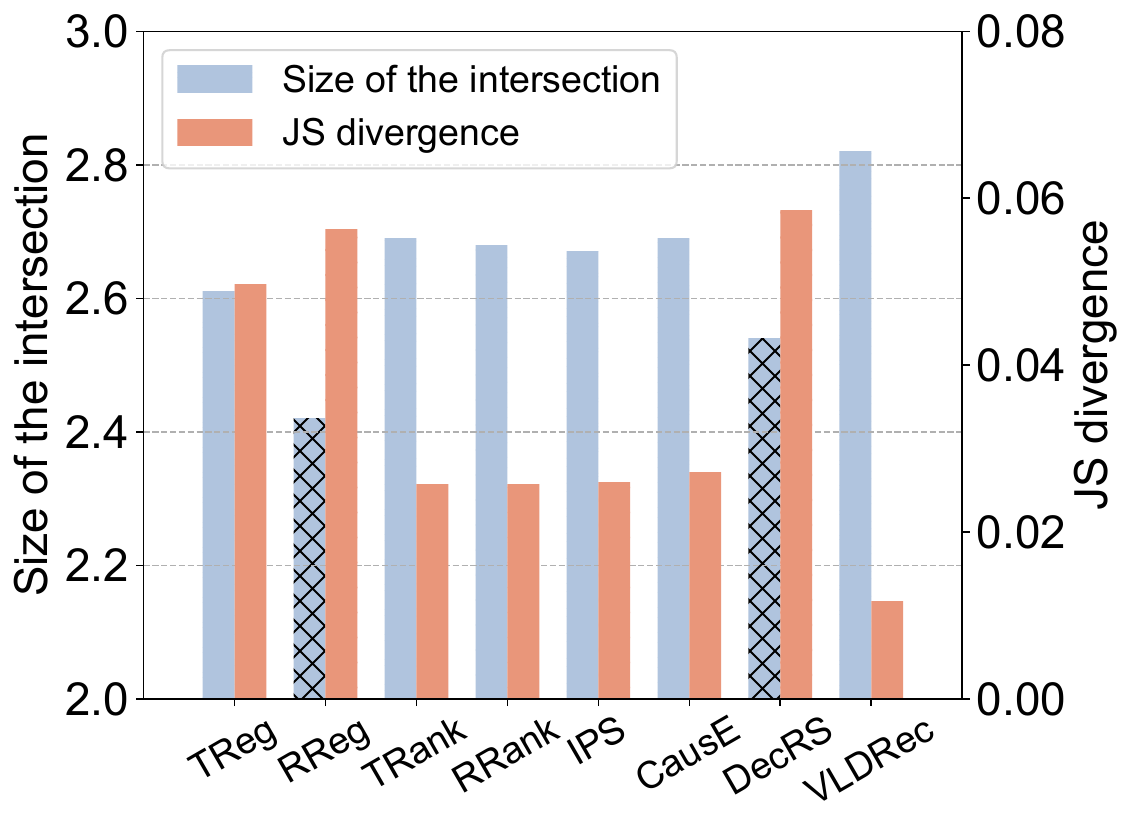}
    \caption{The similarity between the recommendation results of different methods and the ground truth in Wechat.}
    \label{fig:beta_category_study}
\end{figure}

\subsection{Study of Model Capability for Capturing User Preference~(RQ3)}

In terms of the model capability of learning user preferences on micro-videos, besides measuring the overall view time of users, another important dimension is to look at the micro-video content. In other words, the content of recommended micro-videos needs to match the user's interests. 
In this section, we analyze how exactly the recommendation results of different methods match the user’s interests. Specifically, we use the video category information to verify the model's capability of capturing user interests, and this information is only available in Wechat dataset.
Firstly, on an individual level, for each user, we compare the categories of the top five videos in the recommendation result with the categories of the five videos that the user actually watched the longest, and calculate the \textit{size of intersection}. The larger the \textit{size of intersection}, the recommendation results are closer to the user's preference. Results are shown in Fig.~\ref{fig:beta_category_study}, where we can observe that the average \textit{size of intersection} of our VLDRec method is the largest. The RateRegression method performs the worst, and the performance of other methods are also significantly worse than our proposed method. This means that the videos recommended by VLDRec are similar to the user's preference.
Secondly, on a group level, for all users, we aggregate the top five videos recommended by the model and the videos that users actually watched the longest into two collections, respectively. To measure the similarity between these two micro-video distributions, we use the JSD metric. The smaller the JSD value, the higher the similarity between the two distributions. As shown in Fig.~\ref{fig:beta_category_study}, we can observe that VLDRec has the smallest JSD value, while the regression method and DecRS perform poorly overall. 
In a word, VLDRec can effectively match the user's interests, and thus the recommendation results are more similar to the user's historical preference both at the individual and group level, while the common practices in many companies like regression based models only capture the biased preference that are strengthened by video length effect, as a result, its recommendation results cannot effectively match the interests of users.

\section{Conclusion and future work}
\label{sec: conclusion}
In this paper, we aim to tackle the previously untouched problem of video-length effect in recommender systems for online micro-video platforms. We analyze the causes of the video-length effect and propose a VLDRec method for improving micro-video recommendation. By grouping the videos and designing the length-conditioned sampling method, we are able to generate unbiased pairs of training samples and learn the unbiased interests of users through a multi-task learning framework. Experimental results on both public and industrial datasets have proven the superiority of our method over previous solutions in terms of capturing real user preferences in collected user-video view data under a severe video-length bias.

Micro-video recommendation scenario differs from traditional scenarios from sample generation to evaluation due to the different user interfaces of online applications. In this work, we have made some explorations in above areas and we believe that, in the future, more in-depth studies are required, such as how to automatically define the preference labels in a smarter way. More importantly, since it is mainly the change of user interface that has spawned new problems of video-length effect, possible research works related to Computer–Human Interaction~(CHI) seem necessary to expedite the problem resolution. Last but not least, building a new unbiased dataset with randomly exposed videos~\cite{gao2022kuairand} can also help understand user behavior and make better recommendations.

\begin{acks}
This work is supported by the National Key Research and Development Program of China under grant 2020YFA0711403.
This work is also supported by National Natural Science Foundation of China under 62272262, U22B2057, 62171260.
\end{acks}

\bibliographystyle{ACM-Reference-Format}
\bibliography{sample-base}

\end{document}